%% file: barrierlocaltime.tex
\documentclass[numbook, envcountsect, envcountsame, envcountreset, smallextended]{svjour3_My}
\usepackage{amsmath}
\usepackage{amsfonts}
\usepackage{mathptmx}  
\usepackage{hyperref}

\newtheorem{thm}{Theorem}[section]
\newtheorem{lem}[thm]{Lemma}

\newtheorem{prop}[thm]{Proposition}

\newenvironment{pr}{\vspace{1mm}\noindent\textbf{Proof.}}
                   {\vspace{-7mm}\begin{flushright}$\Box$\end{flushright}}

\def\NN{\mathbb{N}}

\def\QQ{\mathbb{Q}}
\def\RR{\mathbb{R}}

\def\EE{\mathbb{E}}

\def\L{\mathcal{L}}

\def\F{\mathcal{F}}
\def\K{\mathcal{K}}

\def\JELname{\textbf{JEL Classification}\enspace}
\def\JEL#1{\par\addvspace\medskipamount{\rightskip=0pt plus1cm
\def\and{\ifhmode\unskip\nobreak\fi\ $\cdot$
}\noindent\JELname\ignorespaces#1\par}}

\title{\textsc{Local time and the pricing of time-dependent barrier options}}

\author{\textsc{Aleksandar Mijatovi\'{c}}
\thanks{I would like to thank Petros Spanoudakis 
for pointing out the problem and for stimulating discussion.
Thanks for many useful comments are due to Dirk Becherer,
Nick Bingham, Johan Tysk and Michalis Zervos.}}

\institute{Aleksandar Mijatovi\'{c},
Department of Mathematics,    
Imperial College London,
Huxley Building, 180 Queen's Gate
London SW7 2AZ, UK,
\email{a.mijatovic@imperial.ac.uk}}

\journalname{}

\begin{document}

\maketitle

\begin{abstract}
A time-dependent double-barrier option is a derivative security that delivers 
the terminal value
$\phi(S_T)$
at expiry 
$T$
if neither of the continuous time-dependent barriers
$b_\pm:[0,T]\rightarrow \RR_+$
have been hit during the time interval
$[0,T]$.
Using a probabilistic approach we obtain a decomposition 
of the barrier option price
into the corresponding European option price minus 
the \textit{barrier premium}
for a wide class of payoff functions
$\phi$,
barrier functions
$b_\pm$
and linear diffusions
$(S_t)_{t\in[0,T]}$.
We show that the barrier premium can be expressed as a sum
of integrals along the barriers
$b_\pm$
of the option's deltas 
$\Delta_\pm:[0,T]\rightarrow\RR$
at the barriers and that
the pair of functions 
$(\Delta_+,\Delta_-)$
solves a system of Volterra integral equations 
of the first kind.
We find a semi-analytic solution for this system in the case of 
constant double barriers and briefly discus a numerical algorithm for the 
time-dependent case.
\end{abstract}


\JEL{G13\and C60}

\markboth{\textsc{Aleksandar Mijatovi\'{c}}}
{\textsc{Local time and the pricing of time-dependent barrier options}}

\section{Introduction}
\label{sec:Intro}
\input{intro.tex}

\section{Integral equations for time-dependent barrier options}
\label{sec:IntegralEq}
\input{Integral.tex}

\section{Examples}
\label{sec:Numerical}
\input{Examples.tex}

\section{Conclusion}
\label{sec:Conclusion}
\input{Conclusion.tex}

\vspace{15mm}

\appendix
\noindent\textbf{\large Appendix}


\section{A change-of-variable formula with local time on curves}
\label{sec:Goran}
\input{changeofvariable.tex}

\section{Analyticity properties of time-dependent barrier option prices}
\label{sec:PDEs}
\input{analyticity.tex}

\section{Proofs of Lemmas~\ref{lem:Boundary},~\ref{lem:regularD} and~\ref{lem:q_Reg}}
\label{sec:regularD}
\input{proof.tex}

\bibliographystyle{plain}
\bibliography{cite}


\end{document}

%% file: intro.tex
Barrier options play an important role in modern financial markets.
They are a less expensive alternative to European options
and trade in large volumes
particularly in foreign exchange. A knock-out double-barrier 
contract is nullified if either of the two barriers
is breached by the underlying asset price process during the life of the
option, and delivers 
$\phi(S_T)$,
for some predefined payoff function
$\phi$,
otherwise.
A knock-in option becomes a European option with payoff
$\phi$
if one of the barriers is hit by the asset price process 
before time
$T$,
and expires worthless otherwise.
Since simple parity relations exist for the prices of knock-in and
knock-out contracts, we shall concentrate only on examining the later. 

The main result of this paper is given by the following representation formula
\begin{eqnarray}
\label{eq:First}
V(0,S_0) =\varphi(0,S_0) - \frac{1}{2}e^{-rT} \left(\int_{0}^{T}\Delta_-(t)q_t(S_0,b_-(t))dt
-\int_{0}^{T}\Delta_+(t)q_t(S_0,b_+(t))dt\right),
\end{eqnarray}
where
$V(0,S_0)$
is the current time-dependent barrier option price, 
$\varphi(0,S_0)$
is the current price of the corresponding 
European payoff and the function
$q_t$
is closely related to the transition density of the process
$S_t$
(see formula~(\ref{eq:KernelFun})),
which is a local volatility process given by the 
stochastic differential equation (SDE)
in~(\ref{eq:stock}).
The functions
$\Delta_-,\Delta_+:[0,T]\rightarrow\RR$
can be interpreted as the limiting values of
the option's deltas
$\frac{\partial V}{\partial S}(t,S_t)$
as the asset price process
$S_t$
approaches either of the two barriers
$b_-(t), b_+(t)$
at time
$t$
(see Theorem~\ref{thm:IntegralEq} 
for more details).
Note that, since the option's payoff is non-negative,
the delta at the lower (resp. upper) barrier is positive
(resp. negative) making the \textit{barrier premium}
in the above formula negative, as one would expect.
By Theorems~\ref{thm:IntegralEq} and~\ref{thm:Final}
the pair of functions
$(\Delta_+,\Delta_-)$
exists and 
solves a system of two Volterra integral equations of the 
first kind, given by~(\ref{eq:DeltaEq}).

An important feature of representation~(\ref{eq:First})
is that it can be used for 
hedging time-dependent barrier options.
Once the system of Volterra integral equations
in~(\ref{eq:DeltaEq})
is solved (numerically
or otherwise), the barrier option price
can be obtained 
by computing the one-dimensional integral in
formula~(\ref{eq:First}),
which, from a numerical point of view, can be done very efficiently. 
Therefore an entire ``spot-ladder'' of option prices
(i.e. a vector of values
$V(0,S_0)$,
where
$S_0$
ranges over a discrete subset in some interval)
can be obtained with little numerical effort,
for
we are only solving the system of Volterra integral equations 
once.
Moreover, since the function
$q_t$
is available in semi-analytic form in most models 
used in practice, 
spot-ladders of deltas and gammas
(i.e. vectors with coordinates
$\frac{\partial V}{\partial S}(0,S_0)$,
$\frac{\partial^2 V}{\partial S^2}(0,S_0)$
respectively, where
$S_0$
takes values in a ``discrete'' interval)
can be found 
by differentiating formula~(\ref{eq:First}),
once we have obtained the solution 
$(\Delta_+,\Delta_-)$
of the 
system in~(\ref{eq:DeltaEq}).
This 
feature of our pricing algorithm
is critical for the risk management of barrier option portfolios
because spot-ladders are one of the most important tools used by traders
for understanding their exposure to adverse 
movements in the underlying market.

Hedging a down-and-out call, when the barrier level is below the strike,
is not dissimilar to hedging the corresponding European option, because 
the presence of the barrier does not destroy the convexity of the payoff
function
and hence the delta
$\frac{\partial V}{\partial S}(t,S_t)$
remains bounded throughout the life of the option.
In the case of a double-barrier knock-out call option the situation is radically
different since the barrier option price is non-convex close to the 
upper barrier at any time 
$t$
before expiry. 
As mentioned earlier,
the value 
$\Delta_+(t)$,
where 
$\Delta_+$
is the function in~(\ref{eq:First}),
is a good approximation for the delta
$\frac{\partial V}{\partial S}(t,S_t)$,
when
$S_t$
is close to the upper barrier, and can hence be used
for hedging. 
We should stress here that
in practice the pair of functions
$(\Delta_+, \Delta_-)$
arises as a solution 
of a non-singular system of linear 
equations (see Subsection~\ref{subsec:Time}
and~\cite{Weiss}, page 179, equation (4.1))
obtained by discretizing the Volterra integral equations
from~(\ref{eq:DeltaEq}),
rather than from a numerical scheme for partial 
differential equations,
where
the first derivative is approximated by 
a difference quotient
that can be unstable close to the boundary
since the value function in that region changes at a very rapid
pace.

Time-dependent barriers arise naturally in financial markets
even if the barriers in the option's contract are constant.
Let
$S_t$
denote the foreign exchange rate and let
the functions
$R_d,R_f:[0,T]\rightarrow\RR$
describe the deterministic term structures of interest rates
in domestic and foreign markets respectively.
The assumption that term structures of interest
rates are deterministic 
is very common 
in the foreign exchange markets 
as
the majority of
barrier option contracts are short dated (with maturities up to one year)
and have little dependence on the stochasticity of interest 
rates.
The forward process
$F_t:=S_t\exp(-t(R_d(t)-R_f(t)))$,
which must be a martingale since it is proportional 
to an asset divided by the domestic bond,
is often modelled directly instead of
the FX rate
$S_t$.
It is clear that the original barrier option's
contract with constant barriers translates into
a contract with time-dependent barriers for 
$F_t$.
Furthermore, by modelling the forward process
directly we can extend 
the representation in~(\ref{eq:First}) 
to the case of time-dependent interest rates.
Note that
the process 
$S_t$
studied in Section~\ref{sec:IntegralEq} (see~(\ref{eq:stock}))
has a constant drift.
But if the functions
$R_d,R_f$
are in
$C^2([0,T])$,
as 
they usually are since market participants
do not like to see kinks in their term structures,
we can model the forward 
$F_t$
by 
$dF_t=F_t\sigma(F_t)dW_t$
and price the equivalent derivative 
with barriers
$b^F_\pm(t):=b_\pm(t)\exp(-t(R_d(t)-R_f(t)))$
using~(\ref{eq:First}). 

A feature frequently encountered in barrier option markets 
is the existence of discontinuous barriers. The barriers are usually
step functions or simply stop being active at a certain time
before expiry.
Since formula~(\ref{eq:First}) works for 
discontinuous 
payoffs, it can be be applied to non-continuous barriers by ``backward integration''.
Time steps would in this case be determined by the intervals of continuity of the barriers.
The procedure starts at the end of the last such interval where the payoff is known and uses
Theorem~\ref{thm:Final} to determine the payoff function at the beginning of that interval.
This produces an equivalent problem with a smaller number of time intervals, so the same procedure
can be reapplied until we obtain the option value at the current time. 

The key idea in behind the proof of Theorem~\ref{thm:IntegralEq}, which yields
representation~(\ref{eq:First}), is in some sense analogous to that used
for finding the integral equation
for the optimal exercise boundary in the American put problem (see Theorem~4.1
in~\cite{Myneni} for a survey account and~\cite{Jacka} for one of the
original derivations).
The smooth-fit principle in the American option problem
implies that the value of the first derivative of the option price at the exercise 
boundary is known, which allows us to obtain a non-linear integral equation for the
exercise boundary by applying It\^{o}'s lemma and taking expectations. 
In the case of barrier options, the boundary of the region 
is specified in advance but the first derivative (i.e. the option's delta)
at the barrier is clearly unknown. 
By judiciously applying Peskir's change-of-variable formula
(see appendix~\ref{sec:Goran} and~\cite{Peskir} for more details)
and taking expectations as in the previous case
it is possible to obtain a Volterra integral
equation of the first kind for the first derivative. 
Unlike with the American put option, where the integral equation
is non-linear in the exercise boundary, 
in the case of barrier options we obtain a linear
equation which,
when discretized,
yields an upper-triangular linear system for the unknown function
that can be solved directly
(see Section~\ref{sec:Numerical} and~\cite{Weiss} for more details). 

The literature on continuously monitored barrier options is vast and varied. It appears 
that two general approaches have been formed.
In the first one, which mainly deals with constant barriers,
one tries to find a path-wise (i.e. robust) hedging strategy with European
derivatives that either uniquely determines or provides an
admissible range for the barrier option price. A model-independent
approach of this kind is exemplified in~\cite{Hobson}. In the case
of the Black-Scholes model, and more generally for models with
symmetric smiles, this approach has been applied to a number
of path-dependent derivatives including constant double-barrier
options (see~\cite{Carr} and~\cite{Carr1}).
The second approach consists of calculating directly the
expectation in a risk-neutral measure of the path-dependent barrier payoff. 
A probabilistic approach using Laplace transforms for constant double-barrier call and put
options in the Black-Scholes model is described in~\cite{GemanYor}.
A method
using the joint density of the stock, its maximum
and its minimum to find the price of time-dependent barrier
options in the Black-Scholes model was pioneered in~\cite{Kunitomo}.
Boundary crossing probabilities for Brownian motion have been used
in~\cite{Novikov} to price single-barrier options where the underlying
asset price process has deterministic time-dependent drift and 
volatility.
In~\cite{Rogers_Barrier} it is shown that
the time-dependent double-barrier option problem 
for geometric Brownian motion can be reduced to the constant 
barriers case by first transforming the state-space and then time.
A static hedge using calls and puts for a 
time-dependent single-barrier option
is described in~\cite{Andersen}. 
The result applies to linear diffusions with compound
Poisson jumps, but the hedging strategy 
depends on knowing the values of the 
barrier contract one is trying to hedge at 
certain times before expiry.
This deficiency  was also noted in~\cite{Jamshidian}
(see page 106), where a simplified derivation
of the main result from~\cite{Andersen}
is given in the case of diffusion processes.
More recent work on
time-dependent double-barrier options for the same kind of asset price
process using analytic tools such as Fourier transforms, Green's
functions and complex integration can be found in~\cite{FourierDoubleBar}, \cite{Paul},
and~\cite{Comment} and~\cite{Pelsser} respectively.
Spectral methods are applied to find constant double-barrier option 
prices in the class of CEV models in~\cite{Linetsky}.
Laplace transforms and Wiener-Hopf factorization are used in~\cite{Pistorius}
to obtain prices and Greeks for constant barrier options where the logarithm
of the underlying asset price process is a
generalized hyper-exponential L\'{e}vy process. This class of processes contains VG, NIG,
CGMY and other models that are of relevance in finance. 
Chapter~12 in~\cite{Lipton} contains a wealth of analytic methods
for pricing a variety of barrier options (time-dependent double barriers
with and without rebate) in specific modelling frameworks (GBM, CEV, Heston)
using the theory of partial differential equations.
A local time approach has been pursued for the study of the static superhedging
of barrier options (see~\cite{Kraft})
and
the decomposition of European options with
convex payoff functions (see~\cite{Carr_Stop}).

In this paper we address the question of pricing
time-dependent single- and double-barrier options where the underlying asset price process
is a linear diffusion with mild regularity conditions on its volatility 
function. 
Our approach is entirely probabilistic, and combines the two approaches
discussed in the previous paragraph. 
We do not make use of 
first passage time distribution, which is prohibitively
complicated in the class of models we are considering.
Instead we employ a  
path-wise analysis of the option price, 
which yields representation~(\ref{eq:First}).
The paper is organised as follows. Section~\ref{sec:IntegralEq}
contains statements and proofs of our main results. In Section~\ref{sec:Numerical}
we propose a semi-analytic solution (using Laplace transforms) of the system of Volterra integral equations
that arises in Theorems~\ref{thm:IntegralEq} and~\ref{thm:Final} in the case of constant double barriers. 
We also discuss discretization methods for the general time-dependent barrier case.
Section~\ref{sec:Conclusion} considers briefly some open questions
related to our results and concludes the paper.

%% file: Integral.tex
In this section our goal is to 
find the integral equations that characterize
the deltas at the barriers and consequently the price
of any time-dependent barrier option.
Before defining precisely the class of exotic options 
we shall consider, let us specify the underlying  
model that provides uncertainty in our economy.
The dynamics of the underlying risky security are given by
a possibly weak solution  (in the sense of Definition 5.5.1 
in~\cite{Karatzas})
of the one-dimensional stochastic differential equation (SDE) 
\begin{eqnarray}
\label{eq:stock}
S_t=S_0+\int_0^t \mu S_u du+\int_0^t S_u\sigma(S_u)dW_u,\quad S_0\in(0,\infty),
\end{eqnarray}
where the function
$\sigma:\RR_+\rightarrow\RR_+$
satisfies
$\sigma(x)>0$
for all 
$x\in(0,\infty)$
and is locally Lipschitz continuous in the 
interval
$(0,\infty)$
(i.e. for any compact set 
$C\subset(0,\infty)$
there exists a positive constant
$K_C$
such that
$|\sigma(x)-\sigma(y)|<K_C|x-y|$
holds
for all 
$x,y\in C$).
These two assumptions 
are the only regularity conditions applied to
the function
$\sigma$
throughout the 
paper.
The constant
$\mu:=r-\delta$
is the risk-neutral drift
given by
the interest rate
$r$
and 
the dividend yield
$\delta$.

The assumptions on 
$\sigma$
imply that the volatility function
$x\mapsto x\sigma(x)$
is also locally Lipschitz continuous
in the interval
$(0,\infty)$
but may vanish at the boundary
$x=0$.
Under these hypotheses Theorem~5.5.15 in~\cite{Karatzas}
yields a filtered probability space
$(\Omega,(\F_t)_{t\in[0,T]},\QQ)$,
with a filtration
$(\F_t)_{t\in[0,T]}$
that satisfies the usual conditions,
and processes 
$S=(S_t)_{t\in[0,T]}$
and
$W=(W_t)_{t\in[0,T]}$
defined on 
$\Omega$,
such that
$W$
is a standard one-dimensional Brownian motion 
with respect to
$(\F_t)_{t\in[0,T]}$
and the process
$S$
solves SDE~\eqref{eq:stock}
up to an explosion time.
Furthermore 
Theorem~5.5.15 guarantees 
uniqueness in law of the 
solution 
$S$.

For some models 
given by SDE~\eqref{eq:stock},
the solution
$S$
can 
reach the boundary point zero of the domain 
$(0,\infty)$
in finite time with strictly positive probability
(e.g. the CEV process, given by 
$\sigma(x)=x^{\rho-1}$,
can reach zero if the parameter
$\rho$
is in the interval
$(0,1)$, 
see~\cite{CEV}).
In such cases the
absorbing boundary condition for the process 
$S$
at 
zero
is  assumed
in Theorem~5.5.15 of~\cite{Karatzas}.
Our aim is to use the measure
$\QQ$
as an equivalent local martingale measure 
for our economy
in the sense 
of~\cite{DS}.
The absorbing boundary condition at zero is therefore 
very natural because any other boundary behaviour would in general
introduce arbitrage (an arbitrage strategy would be to buy the asset 
when it is worth zero and hold it).

The solution 
$S$
of SDE~\eqref{eq:stock}
behaves differently at the other boundary point
of its domain.

\begin{lem}
\label{lem:Boundary}
The process
$S$
does not reach infinity in finite time 
$\QQ$-almost surely.
\end{lem}

For the precise definition of explosion at infinity 
see~\cite{Karatzas}, page 343.
Note that Lemma~\ref{lem:Boundary} implies that
the integrals 
in~\eqref{eq:stock} are defined 
on the entire probability space
$\Omega$
for any fixed time 
$t$,
because the solution process 
$S$
is 
$\QQ$-almost surely finite during the time interval 
$[0,t]$.
For the proof of Lemma~\ref{lem:Boundary} see Appendix~\ref{sec:regularD}.

A continuous time-dependent barrier 
$b:[0,T]\rightarrow (0,\infty) $
is by definition 
a continuous function of finite variation.
In this paper we will mainly be concerned 
with double-barrier options. In order to 
define them we need two
such functions
$b_\pm:[0,T]\rightarrow (0,\infty)$ 
which satisfy 
$b_-(t)<b_+(t)$
for all 
$t\in[0,T]$.
For any fixed time 
$s\in[0,T]$
let the stopping time 
$\tau_s$
be given by
\begin{eqnarray}
\label{eq:stopping_time}
\tau_s:=\inf\{v\in[0,T-s];\>\>S_{s+v}\in\RR_+-(b_-(s+v),b_+(s+v))\>\>\},
\end{eqnarray}
where
$\RR_+:=[0,\infty)$.
Note that by definition we have 
$s+\tau_s\leq T$,
$\{s+\tau_s\leq t\}\in\F_t$
for all 
$t\in[0,T]$,
and
the property 
$s+\tau_s\leq t+\tau_t$
holds for 
$s<t\leq T$.

Let 
$t\in[0,T]$
be the current time.
By definition the fundamental price
$V_t$
of the discounted contract
for
the barrier option 
with a non-negative measurable
payoff function
$\phi:[0,\infty)\rightarrow[0,\infty)$
that started at time
$0$
is given by 
$V_t=\EE\left[
\phi(S_{\tau_0}) I_{\{\tau_0=T\}}\arrowvert \F_t\right],$
where
$I_{\{\tau_0=T\}}$
is the indicator function on
$\Omega$
(see~\cite{Protter_Bubble},
Definition~7).
Since our market is complete,
by Theorem~6 in~\cite{Protter_Bubble} (see also Theorem~3.3 in~~\cite{Cox_Hobson}),
the fundamental price of a derivative security is the smallest initial 
cost of financing a replicating portfolio of that security. 

It was shown in~\cite{Protter_Bubble} (see Subsection~1.5.4)
that the market price of a derivative security equals its fundamental price 
when, in addition to the standard NFLVR assumption of~\cite{DS},
we also stipulate the no-dominance assumption of Merton~\cite{Merton}.
No-dominance intuitively says that, all things being equal,
market participants prefer more to less, and is only violated if
there exists an agent who is willing to buy a dominated security 
at a higher price
(for the mathematical formulation of the no-dominance
assumption see~\cite{Protter_Bubble}, Assumption~3).
No-dominance 
is shown to imply that there are no bubbles 
in the price of the underlying asset or in the price
of a barrier option that is dominated by a call
or a put (see~\cite{Protter_Bubble}, Proposition~1, and Lemma~8 and Theorem~7).
The assumption is consistent with a subclass of models 
given by SDE~\eqref{eq:stock}, namely those that have an equivalent
martingale measure. For example in the CEV framework
($\sigma(x)=x^{\rho-1}$,
where
$\rho\in(0,1]$),
which is known to have a unique equivalent martingale measure
(see~\cite{CEV}),
the no-dominance assumption can be made and the fundamental price 
given by Theorem~\ref{thm:Final}
is the market price.
However the no-dominance assumption cannot be made
when the 
discounted asset price process 
$(\exp(-\mu t)S_t)_{t\in[0,T]}$
is a strict local martingale 
(i.e. there is a bubble in the underlying economy),
which has been shown to be the case for some of the 
models in our framework
(see~\cite{Merton} and~\cite{Tysk}).
In this case the market price of the derivative can 
exceed the fundamental price
given by Theorem~\ref{thm:Final}. 
In other words the market price of the derivative 
is strictly larger than the price of the replicating 
portfolio and little can be said about its dynamics
(see~\cite{Protter_Bubble}, Subsection~1.5.3, Example~5).
It is not an easy task to obtain general 
necessary and sufficient
conditions 
for the existence of an equivalent martingale measure 
for the solution of 
SDE~\eqref{eq:stock},
a topic that merits further research.
In this paper the process 
$V_t$
denotes the fundamental price of the barrier option
in the economy given by~\eqref{eq:stock},
referred to simply as the ``price''
in all that follows.\footnote{Thanks are due to the anonymous 
referee for raising the issue of bubbles and their implications for the
pricing of options.}

Our aim is to find the price 
$V_t$
at any time
$t\in[0,T]$
of a time-dependent double-barrier contract
initiated at time zero.
In order to do this we consider 
the process 
\begin{eqnarray}
\label{eq:Z}
Z_t:=\EE\left[\phi(S_{t+\tau_t}) I_{\{t+\tau_t =T\}}\arrowvert \F_t\right],
\end{eqnarray}
which equals the discounted value of an equivalent
time-dependent barrier contract initiated at 
time
$t$.
Unlike
$V_t$,
which is a 
martingale under the pricing measure 
$\QQ$,
the process
$Z_t$
is not a discounted price process of a security 
in our economy, since at each time 
$t$
it represents the price of a different security,
and hence need not be a martingale (see Lemma~\ref{lem:Z_Sub}).
This is somewhat similar to the well-known observation in interest rate
theory that the short rate (i.e.
the rate at which funds can be borrowed for an infinitesimal 
period of time - also known as the
instantaneous interest rate) corresponds to a different
asset at each time 
$t$
and therefore need not satisfy any no-arbitrage
drift restrictions.
Unlike in the case of the instantaneous interest rate, the drift of
$Z_t$
can be determined uniquely and, as we shall soon see, contains 
all the information needed to obtain the current price of the barrier 
option.
Before exploring some basic properties of the process
$Z$
in the next lemma,
note that definitions~(\ref{eq:stopping_time})
and~(\ref{eq:Z})
also apply to 
single-barrier options
with obvious modifications.

\begin{lem}
\label{lem:Z_Sub}
\begin{description}
\item[(a)]
Let the times
$s,t\in[0,T]$
satisfy
$s\leq t$.
Then
the inequality
$\EE[Z_t\arrowvert\F_s]\geq Z_s$
holds almost surely
in 
$(\Omega,\QQ)$.
If either the upper barrier
$b_+$
is present or the random variable
$\phi(S_T)$
is in 
$L^1(\Omega, \QQ)$,
the process 
$Z_t$
is a non-negative submartingale. 

\item[(b)] 
Assume that the payoff function
$\phi:\RR_+\rightarrow\RR_+$ 
is
continuous 
on the complement of 
a finite set of points 
where it is right-continuous
with left limits, and that,
if
$b_+$
is not present,
the payoff
$\phi(S_T)$
is in
$L^1(\Omega, \QQ).$
Let
the log-normal volatility
$\sigma$
be locally Lipschitz continuous in the interval
$(0,\infty)$
and assume that it satisfies
$\sigma(S)>0$
for all
$S\in(0,\infty)$.
Then the process
$Z_t$
has a continuous modification of the form 
$Z_t=Z(t,S_t)$,
where the continuous function
$Z:[0,T]\times\RR_+\rightarrow\RR_+$
is given by
$Z(t,S):=\EE_{t,S}\left[\phi(S_{t+\tau_t}) I_{\{t+\tau_t=T\}}\right]$.
Let
$C:=\{(t,S)\in[0,T)\times\RR_+;\>b_-(t)<S<b_+(t)\}$,
$B_+:=\{(t,S)\in[0,T)\times\RR_+;\>S>b_+(t)\}$
and
$B_-:=\{(t,S)\in[0,T)\times\RR_+;\>S<b_-(t)\}$
be open subsets of the domain
$[0,T)\times\RR_+$.
Then
$Z$
vanishes on the set
$B_-\cup B_+$,
is of order 
$C^{1,2}(C)$
and satisfies the partial differential equation
$Z_t(t,S)+ \mu S Z_S(t,S)+ \frac{S^2\sigma^2(S)}{2}Z_{SS}(t,S)=0$
for all 
$(t,S)\in C$
with terminal condition
$Z(T,S)  = \phi(S)$ 
for
$S\in (b_-(T),b_+(T))$
and boundary conditions
$Z(t,b_\pm(t)) =  0$ 
for all
$t\in [0,T].$
The same is true for a single-barrier option price
with appropriately modified boundary conditions.
\end{description}
\end{lem}

From now on we shall assume that 
we are working with 
the modification
of the process 
$Z_t$
given in (b)
of Lemma~\ref{lem:Z_Sub}, 
i.e. we will assume that the process 
$Z_t$
is a 
continuous submartingale.
Note also that 
the statement in (a) of Lemma~\ref{lem:Z_Sub} 
is intuitively clear. 
If the underlying asset price 
is between the barriers, the process 
$Z_t$
is a true martingale up to the first time
$S_t$
hits a barrier, because before that stopping
time
$Z_t$
equals 
the discounted barrier option price. 
Since 
$Z_t$
is non-negative, if it were a martingale it would
have to stay at zero from that moment onwards.
But as soon as the stock returns to the 
interval
between the 
barriers, the process 
$Z_t$
assumes again a strictly positive value.
Such behaviour makes its mean drift upwards with time.
We will now give a straightforward but rigorous proof of this fact. 

\begin{pr}
Pick 
$s,t\in[0,T]$
such that 
$s<t$.
Note that 
$s+\tau_s\leq t+\tau_t$
for all paths in 
$\Omega$
and
therefore we have the inclusion
$\{s+\tau_s=T\}\subset\{t+\tau_t=T\}$
and the identity
$I_{\{s+\tau_s=T\}}=I_{\{s+\tau_s>t\}}I_{\{t+\tau_t=T\}}$.
We can now rewrite
$Z_s$,
using the tower property and the fact
$\{s+\tau_s>t\} \in\F_t$,
in the following way
\begin{eqnarray*}
Z_s & = & \EE[\EE[\phi(S_{t+\tau_t})I_{\{s+\tau_s>t\}}I_{\{t+\tau_t=T\}} \arrowvert\F_t]
\arrowvert\F_s] \>\>=
\>\> \EE[Z_t I_{\{\tau_s>t-s\}}\arrowvert\F_s]
 \>\>     \leq  \>\> \EE[Z_t\arrowvert\F_s].
\end{eqnarray*}
The last inequality holds because 
$\phi$,
and hence
$Z_t$,
is non-negative.  
If either of the two integrability conditions 
in 
(a)
of Lemma~\ref{lem:Z_Sub} 
are satisfied,
we get 
$\EE[Z_t]<\infty$
for all
$t\in[0,T]$,
which implies that 
$Z_t$
is a submartingale.
This proves (a).
Part (b) in the lemma is a well-known fact about barrier options.
It suffices to note that the statement (b) is a special case
of Theorem~\ref{thm:analyticity}
in Appendix~\ref{sec:PDEs}.
\end{pr}

Part (b) of Lemma~\ref{lem:Z_Sub} implies
that the partial derivative
$Z_S$
is a continuous function in the open region
$C$,
but the lemma says nothing about the
behaviour of 
$Z_S$
at the boundary
of
$C$.
A key step in obtaining the integral representation for 
the double-barrier option price 
(see equation~(\ref{eq:PriceRep}) in Theorem~\ref{thm:IntegralEq}) 
will be the application 
of Theorem~\ref{thm:Goran}
to the function
$Z:[0,T]\times\RR_+\rightarrow\RR_+$
given in (b) of Lemma~\ref{lem:Z_Sub}.
This step requires a certain regularity of the function
$Z$
and its first derivative 
$Z_S$
close to the boundary of
$C$.
In principle the limit of the delta of the 
double-barrier option price may not exist
as the underlying asset 
$S_t$
approaches the boundary of the region 
$C$.
It does not come as a surprise that 
additional hypotheses on the regularity of the
payoff function
$\phi:\RR_+\rightarrow\RR_+$
as well as of the barriers
$b_\pm:[0,T]\rightarrow\RR_+$
are required for 
the function 
$Z$
to satisfy the assumptions of 
Theorem~\ref{thm:Goran}.
Lemma~\ref{lem:regularD}
gives sufficient conditions
for functions
$\phi$
and
$b_\pm$
that guarantee that the first spatial derivative of the solution
$Z$
of the PDE problem in (b) of Lemma~\ref{lem:Z_Sub}
does not blow up at the boundary 
of the region
$C$.

\begin{lem}
\label{lem:regularD}
Let
the continuous barriers 
$b_\pm:[0,T]\rightarrow\RR_+$
be twice-differentiable and  
assume that the payoff function 
$\phi:[b_-(T),b_+(T)]\rightarrow\RR_+$
satisfies
$\phi(b_-(T))=\phi(b_+(T))=0$
and
is twice-differentiable 
with the second derivative 
$\phi'':(b_-(T),b_+(T))\rightarrow\RR$
which is H\"older continuous 
of order 
$\alpha\in(0,1)$.
If only the lower barrier 
$b_-$
is present,
we additionally assume that the random variable 
$\phi(S_T)$
is in
$L^1(\Omega,\QQ)$.
Then the following holds.
\begin{description}
\item[(a)] The limits
$$\Delta_+(t):=\lim_{\epsilon\searrow0}Z_S(t,b_+(t)-\epsilon)
\>\>\>\>\>\>\mathit{and}\>\>\>\>\>\>\>
\Delta_-(t):=\lim_{\epsilon\searrow0}Z_S(t,b_-(t)+\epsilon)$$
exist for all
$t\in[0,T]$
and are uniform on the interval
$[0,T]$.
\item[(b)] 
For some
$\delta>0$
we have
$\sup_{0<\epsilon<\delta}V(Z(\cdot,b_+(\cdot)-\epsilon))(T)<\infty$
and
$\sup_{0<\epsilon<\delta} V(Z(\cdot,b_-(\cdot)+\epsilon))(T)<\infty$,
where
$V(g)(T)$
denotes the total variation of a function
$g:[0,T]\rightarrow\RR$.
\end{description}
Properties (a) and (b) hold in the time-dependent single-barrier case with obvious modifications.
\end{lem}

Lemma~\ref{lem:regularD} 
is a consequence of Schauder's boundary estimates
for parabolic partial differential equations.
For the proof see Appendix~\ref{sec:regularD}.
Note also that the uniform convergence in the lemma,
together with Theorem~\ref{thm:analyticity},
implies that the delta 
at the barrier 
$\Delta_\pm(t)$
is a continuous function of time for all
$t\in[0,T]$,
if the barrier functions 
$b_\pm$
and the payoff
$\phi$
satisfy the assumptions
in Lemma~\ref{lem:regularD}.
This should be contrasted with the known behaviour of the 
delta
of an up-and-out call option which goes to
minus infinity if, close to expiry, the underlying 
asset approaches the barrier level.

The task now is to understand the path-wise behaviour of the 
process 
$(Z_t)_{t\in[0,T]}$.
For this we will need the important concept of local time.
Recall that the \textit{local time} (at level
$a\in\RR$)
of any continuous semimartingale
$X=(X_t)_{t\in[0,T]}$
on the probability space
$(\Omega,\QQ)$
can be defined as a limit
\label{p:localtime}
$$L_t^a(X):=\lim_{\epsilon\searrow0}\frac{1}{\epsilon}\int_0^tI_{[a,a+\epsilon)}(X_u)
d\langle X,X\rangle_u$$
almost surely in
$\QQ$
(see~\cite{RevuzYor}, page 227, Corollary 1.9),
where 
$\langle X,X\rangle_t$
is the quadratic variation process
as defined in~\cite{RevuzYor}, Chapter IV, Theorem~1.3 and Proposition~1.18.
Notice that this definition can be easily extended to a local time
of 
$X$
along any continuous curve 
$b:[0,T]\rightarrow\RR$
with finite variation by
$L_t^b(X):=L_t^0(X-b)$,
since the process
$X-b$
is still a continuous semimartingale
and the equality 
$\langle X-b,X-b\rangle_t=\langle X,X\rangle_t$
holds
for all
$t$.
For times
$0\leq t < v\leq T$
we denote the local time of
$X$
between 
$t$
and
$v$
by
$L_{t,v}^a(X):=L_v^a(X)-L_t^a(X)$.
It is well-known that the map 
$t\mapsto L_t^a(X)$
is almost surely a non-decreasing continuous function.
With this non-decreasing process one can associate 
a random measure
$dL_t^a(X)$
on the interval
$[0,T]$
the support of which is contained in the set
$\{t\in[0,T];\>\> X_t=a\}$
(see~\cite{RevuzYor}, page 222, Proposition 1.3)

By Lemma~\ref{lem:Z_Sub}
we know that the process
$Z=(Z_t)_{t\in[0,T]}$
is a
non-negative continuous
semimartingale.
Therefore we are at liberty to 
apply the Tanaka formula (see~\cite{RevuzYor}, page 222, 
Theorem~1.2) at level 
$0$
to the non-negative process 
$(Z_t)_{t\in[0,T]}$,
thus obtaining the following path-wise representation
\begin{eqnarray}
\label{eq:path_of_Z}
Z_v & = & Z_t + \int_t^v I_{\{Z_u > 0\}} dZ_u + \frac{1}{2}L_{t,v}^0(Z)\>\>\>\>\>\>\mathrm{for}
\>\>\>\>\>\>0\leq t< v\leq T.
\end{eqnarray}
Using the representation in~(\ref{eq:path_of_Z}) we can prove the following 
proposition, which will play a central role in all that follows.

\begin{prop} 
\label{prop:Pricing_EQ}
Assume that the payoff function
$\phi$
and the barriers 
$b_\pm$
satisfy the assumptions of Lemma~\ref{lem:regularD}
and
let 
$t,v$
be two elements in the interval
$[0,T]$
such that
$t\leq v$.
If the upper barrier 
$b_+$
is present,
the process 
$(\int_t^v I_{\{Z_u > 0\}} dZ_u)_{v\in[t,T]}$
is a continuous martingale and hence 
representation~(\ref{eq:path_of_Z})
is the Doob-Meyer decomposition of the 
submartingale
$(Z_v)_{v\in[t,T]}$.
The following equality must therefore hold
almost surely
\begin{eqnarray}
\label{eq:Rep}
Z_t=\EE[Z_v\arrowvert\F_t]-\frac{1}{2}\EE[L_{t,v}^0(Z)\arrowvert\F_t].
\end{eqnarray}
Assuming that 
$\phi(S_T)$
is in 
$L^2(\Omega,\QQ)$,
the
representation~\eqref{eq:Rep}
holds also in the case where  only the lower barrier
$b_-$
is present.
\end{prop}

If time
$v$
equals expiry
$T$
and time
$t$
equals the current time,
the equality in Proposition~\ref{prop:Pricing_EQ} 
yields a representation of the barrier
option price at 
$t$
as a sum of the current value of the European
payoff
$Z_T$
and the expectation of the local time from now until
expiry. The former quantity is usually available in most models
in a semi-analytic closed form and the latter will be obtained
in Theorem~\ref{thm:IntegralEq} 
by applying the change-of-variable formula from~\cite{Peskir}. 
Note also that intuitively the stochastic integral
$\int_t^v I_{\{Z_u>0\}} dZ_u$
is a martingale because the inegrator
$Z_u$
equals a discounted double-barrier option price on the set
$\{Z_u>0\}$,
which is a martingale.

\begin{pr}
Let
$C$
denote the domain between the barriers
as defined in (b) of Lemma~\ref{lem:Z_Sub}.
Recall that 
$Z_v=Z(v,S_v)$
where the function 
$Z:C\rightarrow \RR_+$
is the solution of the PDE
in (b) of Lemma~\ref{lem:Z_Sub}.
By Lemma~\ref{lem:regularD} we are at liberty to 
apply Theorem~\ref{thm:Goran} 
to the function
$Z$.
In differential form we obtain
$$dZ_u = I_{\{b_-(u)<S_u<b_+(u)\}} Z_S(u,S_u)S_u\sigma(S_u)dW_u +
         \frac{1}{2}\left(I_{\{S_u=b_-(u)\}} \Delta_-(u)dL_u^{b_-}(S)-I_{\{S_u=b_+(u)\}} \Delta_+(u)dL_u^{b_+}(S)\right),$$
where
$Z_S$
denotes the first derivative of 
$Z$
with respect to 
$S$.
The inclusion
$\{Z_u>0\}\subseteq \{b_-(u)<S_u<b_+(u)\}$,
for all
$u\in[0,T]$,
follows from definition~(\ref{eq:Z})
and therefore implies
\begin{eqnarray}
\label{eq:StochInt}
\int_t^v I_{\{Z_u>0\}} dZ_u = \int_t^v I_{\{Z_u>0\}} Z_S(u,S_u)S_u\sigma(S_u)dW_u.
\end{eqnarray}
The function
$Z_S$
is bounded on the domain 
$C$
by Lemma~\ref{lem:regularD},
which implies that the stochastic integral on the right-hand side is
a continuous martingale starting at zero.
By taking expectation on both sides of
equality~(\ref{eq:path_of_Z}) 
we conclude the proof in the double-barrier case.
However,
the same argument can be applied if only the upper barrier
$b_+$
is present. This is because the integrand 
on the right-hand side of~\eqref{eq:StochInt}
is still bounded, making the stochastic integral
in~\eqref{eq:StochInt}
a true
martingale.

In the single-barrier case with only 
$b_-$
present, the argument above does not work because the 
integrand in~\eqref{eq:StochInt} is no longer necessarily bounded.
However the same reasoning shows that the identity in~\eqref{eq:Rep} holds for the stopped 
process 
$Z_{v\wedge\tau_n}=Z(v\wedge\tau_n,S_{v\wedge\tau_n})$,
where the stopping time 
$\tau_n$
is the first passage time of the diffusion 
$S$
into the interval
$[n,\infty)$
after time
$t$,
because the integrand in~\eqref{eq:StochInt}
is bounded.
Jensen's inequality for conditional expectations and the
definition of the process 
$Z$
given in~\eqref{eq:Z}
imply the inequality
$$ \max\{\EE[Z_v^2|\F_t], \EE[Z_{v\wedge\tau_n}^2|\F_t]\}\leq\EE[\phi(S_T)^2|\F_t].$$
Our assumption on 
$\phi(S_T)$
implies
that
$Z_v$
and
$Z_{v\wedge\tau_n}$
are elements of the space
$L^2(\Omega,\QQ)$
for all large natural numbers 
$n$.

By Lemma~\ref{lem:Boundary}
we have that 
$\lim_{n\rightarrow\infty}\tau_n$
is infinite almost surely in
$\Omega$. 
In other words we have almost sure path-wise convergence
$\lim_{n\rightarrow\infty}Z_{v\wedge\tau_n}=Z_v$.
The Cauchy-Schwartz inequality implies the following
\begin{eqnarray*}
\EE[|Z_v-Z_{v\wedge\tau_n}||\F_t] & = & \EE[I_{\tau_n<v}|Z_v-Z_{v\wedge\tau_n}||\F_t] \\
                                   & \leq & \EE[I_{\tau_n<v}|\F_t]^{1/2} \EE[|Z_v-Z_{v\wedge\tau_n}|^2|\F_t]^{1/2}  \\
                                   & \leq & 2\EE[I_{\tau_n<v}|\F_t]^{1/2} \EE[\phi(S_T)^2|\F_t]^{1/2}.  
\end{eqnarray*}
Since the sequence 
$\EE[I_{\tau_n<v}|\F_t]$
converges to zero
$\QQ$-almost surely
as 
$n$
goes to infinity,
we obtain 
$$ 
\EE[Z_v|\F_t]=\lim_{n\rightarrow\infty}\EE[Z_{v\wedge\tau_n}|\F_t]=Z_t-\frac{1}{2}\lim_{n\rightarrow\infty}\EE[L_{t,v\wedge\tau_n}^0(Z)|\F_t]
=Z_t-\frac{1}{2}\EE[L_{t,v}^0(Z)|\F_t],
$$
where the last equality follows by the monotone convergence theorem.
This concludes the proof of the proposition.
\end{pr}

Before we proceed to our main theorem recall that,
for any point
$x\in(0,\infty)$
and time
$t\in(0,T]$,
the density 
$p(t;x,\cdot):(0,\infty)\rightarrow \RR_+$
of the transition function
of the underlying asset price process
$S$,
given by the SDE in~(\ref{eq:stock}),
is characterised by the identity
$\QQ_{x}(S_t\in A)=\int_A p(t;x,y) dy$,
where 
$A$
is any measurable set
in
$(0,\infty)$.
The function 
$p(t;x,\cdot):(0,\infty)\rightarrow \RR_+$
is non-negative but does not necessarily 
integrate to one because the process can
reach zero (and stay there) in finite time.
The existence of 
$p(t;x,y)$
can be deduced from~\cite{ItoMcKean},
Section 4.11,
where it is shown that the transition function of
a diffusion is absolutely continuous with respect to
the speed measure
(see~\cite{ItoMcKean}, page 107, for the definition
of the speed measure). In the case of the process 
$S$
given by~(\ref{eq:stock}),
the speed measure 
is absolutely continuous with respect to the
Lebesgue measure on the interval
$(0,\infty)$
and hence the existence follows.
Furthermore it is proved 
in~\cite{ItoMcKean}
(page 149)
that the function
$p(\cdot;\cdot,y):(0,\infty)\times(0,T]\rightarrow \RR_+$
satisfies the parabolic PDE\label{page:PDEq} in (b) of Lemma~\ref{lem:Z_Sub}
for any 
$y\in(0,\infty)$.
This fact will play a crucial part in the proof of Theorem~\ref{thm:IntegralEq}
(cf. proof of Lemma~\eqref{lem:q_Reg}).
Note also that 
sufficient conditions 
for the existence of densities of solutions of one-dimensional 
SDEs,
which are jointly smooth in all three variables,
are
given in~\cite{Rogers_Density}.
This stronger result
requires
a volatility function that is uniformly bounded away from zero
and is therefore not suited to our purpose.  
For most models that are of relevance in mathematical finance 
the densities can be obtained either in semi-analytic
closed form (see for example~\eqref{eq:density}
and~\eqref{eq:CEV_density})
or numerically. 

The kernels of integral operators 
appearing in Theorems~\ref{thm:IntegralEq}
and~\ref{thm:Final} are related to the transition function
of the asset price process
$(S_t)_{t\in[0,T]}$
and will now be specified precisely.
The quadratic variation
$\langle S, S\rangle_t$
is a continuous non-decreasing adapted process
and as such defines, for each path of
$S_t$,
a measure 
$d\langle S, S\rangle_t$
on the interval
$[0,T]$.
Since the asset price process
$S_t$
is a solution of the SDE in~(\ref{eq:stock}),
this measure is absolutely continuous with respect to
the Lebesgue measure on 
$[0,T]$
and the Radon-Nikodym derivative is given by
$d\langle S,S\rangle_t = S_t^2\sigma(S_t)^2 dt$.
The function
$q_t(x,y)$
that appears in the kernel of the integral operators
in Theorems~\ref{thm:IntegralEq} and~\ref{thm:Final} 
can be defined as
\begin{eqnarray}
\label{eq:KernelFun}
q_t(x,y) & := & p(t;x,y)\frac{d\langle S,S\rangle_t }{dt}\bigg\arrowvert_{S_t=y},
\end{eqnarray}
where
$p(t;x,y)$
is the density defined above.
In the case of the 
geometric Brownian motion 
we have the 
formula
\begin{eqnarray}
\label{eq:density}
q_t(x,y)=\frac{y\sigma}{\sqrt{2\pi t}}\exp\left(-\frac{\left(\log(y/x)-(\mu-\sigma^2/2)t\right)^2}{2\sigma^2t}\right),
\end{eqnarray}
where the drift equals
$\mu = r-d$
and 
$\sigma^2$
is the constant variance.
The function
$p(t;x,\cdot):(0,\infty)\rightarrow \RR_+$
in the case of GBM is a true 
probability density function because the process cannot reach
zero. In the case of the CEV model, given by~(\ref{eq:stock}) 
with absorbing boundary condition at zero
and the log-normal volatility function
$\sigma(x)=\sigma_0x^{\rho-1}$
where
$\rho\in(0,1)$
and
$\sigma_0\in(0,\infty)$,
we have the following closed form expression for the function
$q_t$:
\begin{eqnarray}
\label{eq:CEV_density}
q_t(x,y) & = & 2\sigma_0^2y^{2\rho}(1-\rho)k^{1/(2-2\rho)}(XY^{1-4\rho})^{1/(4-4\rho)}\exp(-X-Y)I_{1/(2-2\rho)}\left(2\sqrt{XY}\right).
\end{eqnarray}
This expression is a consequence of~\eqref{eq:KernelFun}
and the formula for the transition density 
$p_t$,
which can for example be obtained from Theorem~3.5
in~\cite{CEV}.
The function 
$z\mapsto I_\alpha(z)$
is the modified Bessel function of the first kind of order 
$\alpha$
and 
the parameters in~\eqref{eq:CEV_density} are given by 
\begin{eqnarray*}
k & := & \frac{2\mu}{2\sigma_0^2(1-\rho) (\exp(2t\mu(1-\rho)) -1)}, \\
X & := & kx^{2(1-\rho)} \exp(2t\mu(1-\rho)), \\
Y & := &  ky^{2(1-\rho)},
\end{eqnarray*}
where
$\mu$
is the drift in SDE~(\ref{eq:stock}).  We now state one of our main theorems.

\begin{thm}
\label{thm:IntegralEq}
Let 
$S_t$
be the underlying process
given by~(\ref{eq:stock})
and let
$Z_t=Z(t,S_t)$
be the discounted price of a time-dependent single- or double-barrier option contract, starting at 
the current time
$t$,
given in~(\ref{eq:Z}).
Assume further that the 
barriers 
$b_\pm:[0,T]\rightarrow\RR_+$
and the payoff
$\phi:\RR_+\rightarrow\RR_+$
satisfy
the assumptions of Lemma~\ref{lem:regularD}
and that the local volatility function
$x\mapsto\sigma(x)$, $x\in\RR_+$,
satisfies the assumptions in (b)
of Lemma~\ref{lem:Z_Sub}.
In the case where only the lower barrier
$b_-$
is present, we assume in addition that
the variable
$\phi(S_T)$
is in
$L^2(\Omega,\QQ)$.
Let
$\varphi(t,x):=\EE_{t,x}[\phi(S_T)]$
denote the discounted current price of the European contract starting
at time 
$t$,
conditional upon the asset price 
$S_t$
being at level
$x$,
and let the function
$q_t(x,y)$
be as in~(\ref{eq:KernelFun}).
Then the following integral representation for the time-dependent double-barrier option price holds
\begin{eqnarray}
\label{eq:PriceRep}
Z(0,S_0)=\varphi(0,S_0)-\frac{1}{2}\int_{0}^{T}\Delta_-(t)q_t(S_0,b_-(t))dt
                         +\frac{1}{2}\int_{0}^{T}\Delta_+(t)q_t(S_0,b_+(t))dt,
\end{eqnarray}
where 
$\Delta_\pm(t)$
is the limiting value of the delta of the double-barrier option price at 
$b_\pm(t)$
as defined in 
(a) of Lemma~\ref{lem:regularD}.
Furthermore the continuous functions
$\Delta_+, \Delta_-:[0,T]\rightarrow\RR$
satisfy the following linear system of two Volterra integral
equations of the first kind
\begin{eqnarray}
\label{eq:DeltaEq}
\begin{pmatrix}
\varphi(t,b_+(t)) \\
\varphi(t,b_-(t)) \\
\end{pmatrix}
& = &
\frac{1}{2}
\int_t^T Q(t,u)
\begin{pmatrix}
\Delta_+(u) \\
\Delta_-(u) \\
\end{pmatrix}du, 
\end{eqnarray}
where the matrix
$Q(t,u)$,
for
$0\leq t<u\leq T$,
is given by
\begin{eqnarray}
\label{eq:matrix}
Q(t,u)
& := &
\begin{pmatrix}
-q_{u-t}(b_+(t),b_+(u)) & q_{u-t}(b_+(t),b_-(u)) \\
-q_{u-t}(b_-(t),b_+(u)) & q_{u-t}(b_-(t),b_-(u)) \\
\end{pmatrix}.
\end{eqnarray}
In a time-dependent up-and-out (resp. down-and-out) single-barrier case, 
representation~(\ref{eq:PriceRep}) 
contains a single integral along
$b_+$
(resp.
$b_-$).
The integral equation 
that determines the function
$\Delta_+$
(resp.
$\Delta_-$)
in the up-and-out (resp. down-and-out)
case
takes the form of the 
Volterra equation of the first kind
with 
$\pm$ 
equal to $+$ (resp. $-$):
\begin{eqnarray}
\label{eq:OneBar}
\varphi(t,b_\pm(t))\pm\frac{1}{2}\int_t^Tq_{u-t}(b_\pm(t),b_\pm(u))\Delta_\pm(u) du=0.
\end{eqnarray}
\end{thm}

Theorem~\ref{thm:IntegralEq}
yields an integral representation for the double-barrier option price
for a wide variety of local volatility models,
any pair of time-dependent barriers and any payoff function that satisfy 
the assumptions in
Lemma~\ref{lem:regularD}.
Rather surprisingly, knowing the values of the delta at the barriers
for all future times, as well as 
the current price of the corresponding European derivative
(recall that 
$\phi(b_-(T)) =\phi(b_+(T))=0$
for payoffs 
$\phi$
satisfying the assumptions in Lemma~\ref{lem:regularD}),
is enough to obtain the current value of the time-dependent barrier
option. 
Note also that both integrals in equation~(\ref{eq:PriceRep}) are negative since
$\Delta_-(t)>0$
(resp.
$\Delta_+(t)<0$),
which intuitively follows from the fact that 
the barrier option price is
increasing (resp. decreasing) as the asset price
moves away from (resp. approaches) the lower (resp. upper)
barrier. As expected this makes the barrier option cheaper
than its European counterpart.
Representation~(\ref{eq:PriceRep}) 
therefore decomposes the double-barrier option price
into the European option price and the \textit{barrier premium}.

In order to include the payoff functions 
$\phi$
that are of interest in applications 
(e.g. the up-and-out call option payoff
$\phi(S)=(S-K)^+I_{(0,b_+(T))}(S)$
or the payoff of a double-no-touch
$\phi(S)=I_{(b_-(T),b_+(T))}(S)$),
we must relax the smoothness requirements 
for the function
$\phi$
stipulated in 
Lemma~\ref{lem:regularD}.
This will be done in 
Theorem~\ref{thm:Final},
where we will show that the integral representation for the 
price~(\ref{eq:PriceRep})  
and the integral
equation for functions
$\Delta_\pm$~(\ref{eq:DeltaEq})
continue to hold.

Before proceeding to the proof of Theorem~\ref{thm:IntegralEq}
we need the following lemma that bounds the growth of the 
function 
$q$,
defined in~\ref{eq:KernelFun},
over short time intervals.

\begin{lem}
\label{lem:q_Reg}
Let 
$K$
be a compact interval contained in 
$(0,\infty)$.
Then there exists a positive constant 
$C_K$
such that the inequality
$$q_{u-t}(x,y)<\frac{C_K}{\sqrt{u-t}} $$
holds
for all 
$t<u\leq T$
and 
$x,y\in K$.
\end{lem}

The proof of Lemma~\ref{lem:q_Reg}
is contained in Appendix~\ref{sec:PDEs}.
Note that the constant
$C_K$
in Lemma~\eqref{lem:q_Reg}
depends only on the compact set 
and the inequality therefore holds uniformly
on
$K$.
If the function 
$\sigma$
in SDE~\eqref{eq:stock} were uniformly bounded away from
zero, the estimate in Lemma~\eqref{lem:q_Reg} would 
hold on the entire domain 
$(0,\infty)$.

Lemma~\eqref{lem:q_Reg} implies that 
integral equations~(\ref{eq:DeltaEq}) and~(\ref{eq:OneBar})
have weakly singular kernels and that the inequalities 
$q_{u-t}(b_\pm(t),b_\pm(u))<\frac{M}{\sqrt{u-t}}$
hold
for all 
$u\in(t,T]$
where 
$M$
is
a positive constant
($\pm$
denotes either 
$+$
or
$-$).
The linear operator in~(\ref{eq:OneBar})
(resp.~(\ref{eq:DeltaEq})) 
is compact on the Banach spaces
of continuous functions 
$C([0,T])$
(resp.
$C([0,T])\times C([0,T])$) 
with the supremum norm,
and as such has 
$0$
in its spectrum. 
Note that by construction 
equations~(\ref{eq:DeltaEq}) and~(\ref{eq:OneBar})
have a continuous solution. 
The uniqueness of this solution is a much more subtle question,
equivalent to asking whether 
$0$
in the spectrum of the operator is an eigenvalue.
Since equations~(\ref{eq:DeltaEq}) and~(\ref{eq:OneBar})
are of the first kind and the Fredholm alternative
(which provides a 
general answer to the question of uniqueness of solutions 
for the integral equations of the second kind) cannot be used,
it is difficult to answer the question in general.
However for a time-dependent single-barrier case in the Black-Scholes model
see Proposition~\ref{prop:Unique}.
Let us now proceed to the proof of Theorem~\ref{thm:IntegralEq}.

\begin{pr}
Let us start by considering a time-dependent double-barrier option.
Let 
$C$
be the domain between the barriers
as defined in (b) of Lemma~\ref{lem:Z_Sub}.
We begin by applying Theorems~\ref{thm:Goran} 
and~\ref{thm:analyticity}	
to the process
$Z_t=Z(t,S_t)$,
where the function 
$Z$
is the solution of the PDE from (b) of Lemma~\ref{lem:Z_Sub}.
For any pair of times
$t,v\in[0,T]$,
such that 
$t<v$,
we therefore obtain the following path-wise representation
$$Z_v=Z(t,S_t)+ \int_t^v I_{\{b_-(u)<S_u<b_+(u)\}} Z_S(u,S_u)S_u\sigma(S_u)dW_u +
         \frac{1}{2} \int_t^v \Delta_-(u)dL_u^{b_-}(S)- \frac{1}{2} \int_t^v  \Delta_+(u)dL_u^{b_+}(S),$$
where the functions 
$\Delta_+$
and
$\Delta_-$
are defined in Lemma~\ref{lem:regularD}.
The random measures
$dL_u^{b_\pm}(S)$
are by definition equal to the 
well-defined random measures
$dL_u^0(S-b_\pm)$
and the functions
$\Delta_\pm$
are continuous by Theorem~\ref{thm:analyticity}	and 
(a) of Lemma~\ref{lem:regularD}
and are hence Borel measurable.
Since the function
$Z_S:C\rightarrow\RR_+$
is bounded, this equality yields a Doob-Meyer decomposition of the
submartingale 
$(Z_v)_{v\in[t,T]}$.
Since such a decomposition is unique, 
Proposition~\ref{prop:Pricing_EQ}
implies the following identity for the finite variation
processes
\begin{eqnarray}
\label{eq:LT}
L_{t,v}^0(Z)=\frac{1}{2}\int_t^v\Delta_-(u)dL_u^{b_-}(S)-
\frac{1}{2}\int_t^v\Delta_+(u)dL_u^{b_+}(S).
\end{eqnarray}

The main idea for the proof 
of Theorem~\ref{thm:IntegralEq}
is to use 
the equality 
in Proposition~\ref{prop:Pricing_EQ}
to obtain the representation of the option price and the 
integral equations in the theorem. We must therefore 
find the expectation 
$\EE_{t,S_t}[L_{t,v}^0(Z)]$
using identity~(\ref{eq:LT}).
Let us start by proving the following.

\noindent \textbf{Claim.} For any continuous function
$f:[0,T]\rightarrow \RR$
of finite variation and for 
all
$t,v\in[0,T]$,
such that
$t<v$,
the equality 
$$\EE_{t,S_t}\left[\int_t^vf(u)dL_u^{b_\pm}(S)\right] =\int_t^vf(u)q_{u-t}(S_t,b_\pm(u))du$$
holds,
where
$q_{u-t}(x,y)$
is given in~(\ref{eq:KernelFun}).

Recall that, since the process
$(S-b_\pm)_{t\in[0,T]}$
is a continuous semimartingale,
there exists a modification of the local time
$L_{t,v}^a(S-b_\pm)$
such that the map
$a\mapsto L_{t,v}^a(S-b_\pm)$
is right-continuous and has 
left limits
for every 
$v\in[t,T]$
almost surely in
$\Omega$.
The function is therefore 
Lebesgue measurable and the occupation times formula
(see~\cite{RevuzYor}, Chapter VI, Corollary~1.6)
implies
$\int_t^v I_{[0,\epsilon)}(S_u-b_\pm(u))d\langle S,S\rangle_u = 
\int_0^\epsilon L_{t,v}^a(S-b_\pm)da$
for
$\epsilon>0$.
By taking expectations and dividing by
$\epsilon$
on both sides of this equality we obtain
\begin{eqnarray}
\label{eq:OccTime}
\int_t^v \frac{1}{\epsilon} \EE_{t,S_t} \left[I_{[0,\epsilon)}(S_u-b_\pm(u))S_u^2\sigma(S_u)^2\right]du = 
\frac{1}{\epsilon} \int_0^\epsilon \EE_{t,S_t}\left[L_{t,v}^a(S-b_\pm)\right] da.
\end{eqnarray}
The integrand on the left equals
$\frac{1}{\epsilon}\int_{b_\pm(u) }^{b_\pm(u)+\epsilon}q_{u-t}(S_t,y)dy$
and
in the limit 
$\epsilon\searrow0$
we obtain
$ q_{u-t}(S_t,b_\pm(u))$
for all
$u\in(t,v]$.
Since for all small 
$\epsilon$
we have the inequality
$q_{u-t}(S_t,y)<\frac{M}{\sqrt{u-t}}$
for some constant
$M$
and
$y\in[b_\pm(u),b_\pm(u)+\epsilon]$
(see Lemma~\ref{lem:q_Reg}),
we can apply the bounded convergence theorem to the left-hand
side of~(\ref{eq:OccTime})
to obtain 
$\int_t^vq_{u-t}(S_t,b_\pm(u))du$
for all values 
$S_t$
(including
$S_t=b_\pm(t)$).

The right-hand side
of~(\ref{eq:OccTime}) will
converge to 
$\EE_{t,S_t}\left[L_{t,v}^0(S-b_\pm)\right]$
by the fundamental theorem of calculus,
if we can show that the function
$a\mapsto \EE_{t,S_t}\left[L_{t,v}^a(S-b_\pm)\right] $
is continuous at
$a=0$.
Tanaka's formula yields the following representation for local time
$$\frac{1}{2}L_{t,v}^a(S-b_\pm) = \Psi_{t,v}(a) + \int_t^vI_{\{S_u\leq b_\pm(u)+a\}}S_u\sigma(S_u)dW_u+
\int_t^v I_{\{S_u\leq b_\pm(u)+a\}}(\mu S_u-b_\pm'(u))du,$$
where
$\Psi_{t,v}(a) := (a-(S_v-b_\pm(v)))^+-(a-(S_t-b_\pm(t)))^+$
and, as usual,
$(x)^+:=\max\{x,0\}$
for any
$x\in\RR$
(see~\cite{RevuzYor}, Chapter VI, Theorem~1.2).
Note that the variable
$\Psi_{t,v}(a)$
is Lipschitz continuous in
$a$
with a Lipschitz constant equal to
$1$
for all elements in
$\Omega$.
By taking expectation on both sides we find
\begin{eqnarray}
\label{eq:FinalExp}
\EE_{t,S_t}\left[L_{t,v}^a(S-b_\pm)\right]=2 \EE_{t,S_t}[\Psi_{t,v}(a)] +2\EE_{t,S_t}\left[
\int_t^v I_{\{S_u\leq b_\pm(u)+a\}}(\mu S_u-b_\pm'(u))du
\right],
\end{eqnarray}
since the integrand in the stochastic integral is bounded
and hence 
the martingale term vanishes in expectation.
The quantity 
$\EE_{t,S_t}[\Psi_{t,v}(a)]$
is continuous in
$a$,
while the second expectation on the right-hand side can be rewritten
using Fubini's theorem in the following way
$\int_t^v F(a,u) du$,
where the function 
$F$
is given by
$F(a,u):=C_0(u)+\int_0^{b_\pm(u)+a}(\mu y-b_\pm'(u))p(u-t;S_t,y)dy$,
the function
$p$
denotes the density of the asset price process
$S$
in the interval
$(0,\infty)$
and 
$C_0(u):=-b_\pm(u)\QQ_{S_t}(S_u=0)$
is a function independent of
$a$.
The 
estimate 
$p(u-t;x,y)\leq \frac{C}{\sqrt{u-t}}$,
for a positive constant
$C$
independent of
$x,y$
in a compact subset of
$(0,\infty)$
(see Lemma~\ref{lem:q_Reg}),
implies that the function
$a\mapsto F(a,u)$
possesses a partial derivative that is bounded 
in the following way
$\lvert\frac{\partial F}{\partial a}(a,u)\rvert\leq \frac{D}{\sqrt{u-t}}$
for all 
$u\in[t,v]$.
$D$
is some positive constant
independent of
$S_t$.
Lagrange's theorem now
implies that the integral
$\int_t^v F(a,u) du$
is a continuous function of
$a$.
We have therefore shown that~(\ref{eq:FinalExp})
is continuous in 
$a$
and hence proved that the key identity
$$\EE_{t,S_t}[L_{t,v}^{b_\pm}(S)]=\int_t^vq_{u-t}(S_t,b_\pm(u))du$$
follows from~(\ref{eq:OccTime})
upon taking the limit
$\epsilon\searrow0$.

For every path 
$\omega$
in the probability space
$\Omega$,
the function
$v\mapsto L_{t,v}^{b_\pm}(S)(\omega)$
has finite variation and is continuous.
Since the same is true of the function
$f$
in our claim,
we can use the integration by parts formula to obtain
the equality
$\int_t^vf(u)dL_u^{b_\pm}(S) =f(v)L_{t,v}^{b_\pm}(S)-
\int_t^vL_{t,u}^{b_\pm}(S) df_u$,
where
$df_u$
is the Radon measure on the interval
$[t,v]$
induced by 
$f$.
By taking expectations on both sides of this identity 
and applying Fubini's theorem to the integral on the 
right, which is justified since local time is a non-negative
function, we obtain the following sequence of equalities:
\begin{eqnarray*}
\EE_{t,S_t} \left[\int_t^vf(u)dL_u^{b_\pm}(S)\right] & = &
f(v)\EE_{t,S_t}[L_{t,v}^{b_\pm}(S)] -\int_t^v\EE_{t,S_t}[L_{t,u}^{b_\pm}(S)] df_u \\
& = & f(v)\EE_{t,S_t}[L_{t,v}^{b_\pm}(S)] - \int_t^vdf_u\int_t^uq_{s-t}(S_t,b_\pm(s))ds \\
& = & f(v)\EE_{t,S_t}[L_{t,v}^{b_\pm}(S)] - \int_t^v(f(v)-f(s))q_{s-t}(S_t,b_\pm(s))ds \\
& = & \int_t^vf(s)q_{s-t}(S_t,b_\pm(s))ds. 
\end{eqnarray*} 
The third equality follows by Fubini's theorem and the last one is 
a consequence of the formula for the expectation of local time.
This proves the claim.

In order to apply the claim to the identity in~(\ref{eq:LT}),
we need to approximate the continuous functions
$\Delta_\pm$
on the interval
$[t,v]$
by sequences of uniformly bounded continuous functions
$(f_n^\pm:[t,v]\rightarrow\RR)_{n\in\NN}$
with finite total variation (since the functions
$\Delta_\pm$
are bounded on
$[t,v]$,
we can take piecewise linear approximations on a uniform 
grid in
$[t,v]$).
For each path
$\omega\in\Omega$
the dominated convergence theorem implies that the equality
$\int_t^v\Delta_\pm(u)dL_u^{b_\pm}(S)(\omega)=
\lim_{n\rightarrow\infty}\int_t^vf_n^\pm(u)dL_u^{b_\pm}(S)(\omega)$
holds.
Since the functions
$f_n^\pm$
are uniformly bounded by some constant
$K$,
the random variables
$|\int_t^vf_n^\pm(u)dL_u^{b_\pm}(S)|$
are bounded by 
$KL_t^{b_\pm}(S)$,
which is an integrable random variable.
Another application of
the dominated convergence theorem and the above claim therefore yield the following 
equalities
$$\EE_{t,S_t}\left[ \int_t^v\Delta_\pm(u)dL_u^{b_\pm}(S)\right]=
\lim_{n\rightarrow\infty} \EE_{t,S_t}\left[\int_t^vf_n^\pm(u)dL_u^{b_\pm}(S)\right]
=\int_t^v\Delta_\pm(u)q_{u-t}(S_t,b_\pm(u))du,$$
for any pair of times
$t,v\in[0,T]$
that satisfy
$t<v$.

We can now apply the last equality to equation~(\ref{eq:LT})
to find the expectation of the local time of the time-dependent double-barrier option price.
In other words, by Proposition~\ref{prop:Pricing_EQ}
we have the following representation for the expectation 
of the double-barrier 
option price process
\begin{eqnarray}
\label{eq:FinalRep}
\EE_{t,S_t}[Z_v]=Z(t,S_t)+
\frac{1}{2} \int_t^v\Delta_-(u)q_{u-t}(S_t,b_-(u))du-
\frac{1}{2} \int_t^v\Delta_+(u)q_{u-t}(S_t,b_+(u))du
\end{eqnarray}
for all 
$t,v\in[0,T]$
satisfying
$t<v$
and all values of
$S_t$.
The representation of the 
double-barrier option price~(\ref{eq:PriceRep})
in the theorem can be obtained by taking
$t=0$
and
$v=T$
in equation~(\ref{eq:FinalRep}).
The system of integral equations~(\ref{eq:DeltaEq}) 
for 
$(\Delta_+, \Delta_-)$
also follows from formula~(\ref{eq:FinalRep})
by taking
$v=T$,
$S_t=b_+(t)$
and
$S_t=b_-(t)$,
and observing that 
$Z(t,b_-(t))=Z(t,b_+(t))=0$
for all
$t\in[0,T],$
since the double-barrier contract that starts 
at the barrier is worth zero by definition.
This completes the proof of the double-barrier case.
The single-barrier case can be obtained by making straightforward 
modifications to the preceding proof.
\end{pr}

Our next task is to relax the assumptions on the smoothness of 
the payoff function
$\phi:\RR_+\rightarrow\RR_+$
made in Theorem~\ref{thm:IntegralEq}.
This is crucial, as we would like to be able to apply our methodology
to the payoffs that arise in practice, such as the up-and-out call option
payoff
$\phi(S)=(S-K)^+I_{(0,b_+(T))}(S)$
or the payoff of a double-no-touch
$\phi(S)=I_{(b_-(T),b_+(T))}(S)$.
The following theorem allows us to do precisely that.

\begin{thm}
\label{thm:Final}
Let
$\phi:\RR_+\rightarrow\RR_+$
be a payoff function
that is continuous everywhere except at a
finite set of points where it is 
right-continuous and has
left limits.
Assume further that the 
barriers 
$b_\pm:[0,T]\rightarrow\RR_+$
satisfy
the assumptions of Lemma~\ref{lem:regularD}
and that the asset price process
$(S_t)_{t\in[0,T]}$
is
given by~(\ref{eq:stock}).
In the case where only the lower barrier
$b_-$
is present, we assume in addition that
the variable
$\phi(S_T)$
is in
$L^2(\Omega,\QQ)$.
Let
$Z_t=Z(t,S_t)$
be the discounted price of a time-dependent single- or double-barrier option contract, starting at 
the current time
$t$,
given in~(\ref{eq:Z}),
and let
$\varphi(t,x):=\EE_{t,x}[\phi(S_T)I_{(b_-(T),b_+(T))}(S_T)]$
denote the discounted price of the European contract 
at the current time 
$t$
conditional upon the asset price 
$S_t$
being at level
$x$.
Then there exist 
measurable functions
$\Delta_+, \Delta_-:[0,T]\rightarrow\RR$,
which are
in
$L^1([0,T],m_\pm(dt))$
and are not necessarily continuous or bounded,
such that 
the following integral representation for the double-barrier option price holds
\begin{eqnarray}
\label{eq:PriceRepNonCont}
Z(0,S_0)=\varphi(0,S_0)-\frac{1}{2}\int_{0}^{T}\Delta_-(t)q_t(S_0,b_-(t))dt
                         +\frac{1}{2}\int_{0}^{T}\Delta_+(t)q_t(S_0,b_+(t))dt.
\end{eqnarray}
The measure
$m_\pm$
is absolutely continuous with respect to the 
Lebesgue measure and the Radon-Nikodym derivative
is given by
$\frac{dm_\pm}{dt} = q_t(b_\pm(0),b_\pm (t))$,
where 
$\pm$
is either 
$+$
or
$-$.
Furthermore the functions
$\Delta_+, \Delta_-$
satisfy the linear system of Volterra integral
equations of the first kind
given by~(\ref{eq:DeltaEq}).
In a time-dependent up-and-out (resp. down-and-out) single-barrier case 
there exists a measurable function
$\Delta_+$
(resp.
$\Delta_-$), 
which is contained in
$L^1([0,T],m_+(dt))$
(resp.
$L^1([0,T],m_-(dt))$)
and is
not necessarily continuous or bounded,
such that 
the discounted option price
$Z(0,S_0)$
has the following integral representation
\begin{eqnarray*}
Z(0,S_0)=\varphi(0,S_0)\pm\frac{1}{2}\int_{0}^{T}\Delta_\pm(t)q_t(S_0,b_\pm(t))dt,
\end{eqnarray*}
where 
$\pm$
equals either 
$+$
or
$-$.
The integral equation 
satisfied by the function
$\Delta_+$
(resp.
$\Delta_-$)
takes the form~(\ref{eq:OneBar}).
\end{thm}

At first glance Theorems~\ref{thm:IntegralEq}
and~\ref{thm:Final} look similar. The difference
lies in the fact that Theorem~\ref{thm:Final} applies to
a much wider class of payoff functions
$\phi$
that do not satisfy the hypothesis of Lemma~\ref{lem:regularD}
and furthermore invalidate its conclusions. This makes it impossible 
to apply
the key local time formula from Theorem~\ref{thm:Goran},
which provided the core of the proof of Theorem~\ref{thm:IntegralEq}.
These analytical difficulties will be circumvented by a careful approximation
argument yielding the existence of 
$L^1$
functions
$\Delta_-,\Delta_+$,
which satisfy integral equation~(\ref{eq:DeltaEq})
and give the desired representation for the time-dependent barrier
option price.

As an illustration of the difference between 
Theorems~\ref{thm:Final} 
and~\ref{thm:IntegralEq}
consider the following.
It is well-know that
the delta of a short position in an up-and-out call option becomes 
arbitrarily large
if, close to expiry, the asset price approaches the
barrier. In particular this implies that 
$\Delta_+$
cannot be bounded close to expiry. A trader trying to hedge
this position would have to buy unlimited amounts of the underlying
asset. Since the gamma of the short position in the up-and-out call option
is large and positive close to the barrier, this delta hedge would be
very profitable if the barrier were not touched. However if the barrier
were broken, the large accumulation of the underlying asset would become
a huge problem. This is why, in practice, such a position close to
expiry would be left unhedged. Let us now proceed to the proof of
Theorem~\ref{thm:Final}. 

\begin{pr} 
Let 
$E\subset\RR_+$
be the finite set of discontinuities of the payoff function
$x\mapsto\phi(x)I_{(b_-(T),b_+(T))}(x)$,
which we also denote by 
$\phi$
for notational convenience.
We start by constructing a sequence of functions
$\phi_n:\RR_+\rightarrow\RR_+$
that satisfy the assumptions of Lemma~\ref{lem:regularD}
and have the following two properties
\begin{description}
\item[(1)] $\phi_n(x)\leq\phi_{n+1}(x)$,
for all
$x\in\RR_+$
and
all
$n\in\NN$,
and
\item[(2)] 
$\lim_{n\rightarrow\infty}\phi_n(x)=\phi(x)$,
for all
$x\in\RR_+-E$.
\end{description}
Let
$p,r\in E$
be two consecutive points 
in 
$E$
such that
$p<r$.
In other words the function
$\phi$
is continuous on the interval
$[p,r)$
and has a limit at 
$r$.
By Stone-Weierstrass theorem
for each 
$n\in\NN$
there exists an element
$f_n\in C^3([p,r])$,
such that the inequalities
$\max\{\phi(x)-\frac{1}{n+1},0\} \geq f_n(x)\geq \max\{\phi(x)-\frac{1}{n},0\}$
hold for all
$x\in[p,r)$.
The construction implies that the sequence
$(f_n)_{n\in\NN}$
satisfies property
(1) 
for all
$x\in[p,r]$
and property 
(2)
for all
$x\in[p,r)$.
The complement 
$\RR_+-E$
consists of a finite number of open intervals with the same properties as
$(p,r)$.
For each point
$p\in E$
we can choose a decreasing sequence of open intervals
$N_n^p$
such that
$\{p\} =\cap_{n=1}^\infty N_n^p$,
$E\cap N_n^p=\{p\}$,
for all 
$n\in\NN$,
and
$N_1^p\cap N_1^r=\emptyset$
for any
$r\in E-\{p\}$.
Note that on 
the components of
$\RR_+-E$
adjacent to any
$p\in E$
we have already constructed sequences 
$(f_n)_{n\in\NN}$
and
$(g_n)_{n\in\NN}$
of 
$C^3$ 
functions
that converge to
$\phi$
in the required way.
In the complement 
of the neighbourhood
$N_n^p$,
we define 
$\phi_n(x)$
to equal either 
$f_n(x)$
or
$g_n(x)$,
depending on
$x$
being larger or smaller than 
$p$.
We can now easily extend 
$\phi_n$
to the interval
$N_n^p$
so that the resulting function is
$C^3$
and 
property~(1)
remains true on any neighbourhood of
$p$.
Since 
$\{p\} =\cap_{n=1}^\infty N_n^p$,
property (2)
is also satisfied.

Let 
$Z^n_t=Z^n(t,S_t)$
denote the process given in~(\ref{eq:Z})
that corresponds to the payoff function
$\phi_n$.
It is clear that property (1)
implies the  inequality
$Z^n(t,S) \leq Z^{n+1}(t,S)$
for all points
$(t,S)\in C$,
where the region 
$C$
is defined in (b) of Lemma~\ref{lem:Z_Sub},
and
all
$n\in\NN$.
Since the set
$E$
is finite,
properties (1) and (2) and the monotone convergence
theorem imply the equality
$\lim_{n\rightarrow\infty}Z^n(t,S_t)=Z(t,S_t)$,
where
$Z_t=Z(t,S_t)$
is given by~(\ref{eq:Z}).
Since 
$\phi_n$
satisfies the hypotheses of 
Lemma~\ref{lem:regularD}, expression~(\ref{eq:FinalRep})
can be rewritten as 
\begin{eqnarray}
\label{eq:FinalRepSeq}
\EE_{0,S_0}[\phi_n(S_T)]=Z^n(0,S_0)+
\frac{1}{2} \int_0^T\Delta_-^n(t)q_{t}(S_0,b_-(t))dt-
\frac{1}{2} \int_0^T\Delta_+^n(t)q_{t}(S_0,b_+(t))dt
\end{eqnarray}
for all 
$n\in\NN$
and 
$t,v\in[0,T]$
such that 
$t<v$.
For every
$n\in\NN$
the deltas at the barriers exist by Lemma~\ref{lem:regularD}
and are given by
$\Delta_\pm^n(t):=\lim_{k\rightarrow\infty} Z_S^n(t,b_\pm(t)\mp\epsilon_k)$,
where
$(\epsilon_k)_{k\in\NN}$
is a positive monotonically decreasing sequence accumulating at zero.
Notice that, since
$Z^n(t,b_\pm(t))=0$
for all
$t\in[0,T]$,
Lagrange's theorem implies the equalities
$$ \Delta_+^n(t)=-\lim_{k\rightarrow\infty} \frac{Z^n(t,b_+(t)-\epsilon_k) }{\epsilon_k},
\>\>\>\>\>\>\>\>\>\>\>\>
\Delta_-^n(t)=\lim_{k\rightarrow\infty} \frac{Z^n(t,b_-(t)+\epsilon_k)}{\epsilon_k},
$$
for each
$n\in\NN$.
Since the inequality
$Z^n(t,b_+(t)-\epsilon_k) \leq Z^{n+1}(t,b_+(t)-\epsilon_k)$
(resp. 
$Z^n(t,b_-(t)+\epsilon_k) \leq Z^{n+1}(t,b_-(t)+\epsilon_k)$)
holds for any 
$n\in\NN$
and all
$k\in\NN$,
it follows that 
$0\geq\Delta_+^n(t) \geq \Delta_+^{n+1}(t)$
(resp.
$0\leq\Delta_-^n(t)\leq \Delta_-^{n+1}(t)$)
for all
$t\in[0,T]$.
In other words the negative sequence
$(\Delta_+^n(t))_{n\in\NN}$
(resp. positive sequence
$(\Delta_-^n(t))_{n\in\NN}$)
is decreasing
(resp. increasing)
at any time
$t$
and hence converges to
its infimum (resp. supremum),
which is not necessarily finite.
We can therefore define measurable functions
$\Delta_+:[0,T]\rightarrow [-\infty,0],\>\> \Delta_-:[0,T]\rightarrow [0,\infty]$
in the following way
$\Delta_+(t):=\lim_{n\rightarrow\infty}\Delta^n_+(t)$,
$\Delta_-(t):=\lim_{n\rightarrow\infty}\Delta^n_-(t)$.
By applying monotone convergence theorem
to all the integrals in expression~(\ref{eq:FinalRepSeq})
we obtain formula~(\ref{eq:PriceRepNonCont})
in the theorem.
Furthermore,
formula~(\ref{eq:PriceRepNonCont})
implies that the integrals
$\int_{0}^{T}\Delta_\pm(t)q_t(S_0,b_\pm(t))dt$
are finite. Since the functions
$\Delta_+, \Delta_-$ 
do not change sign,
they clearly define elements in
$L^1([0,T],m_+(dt))$,
$L^1([0,T],m_-(dt))$
respectively.
The system of Volterra integral equations~(\ref{eq:DeltaEq})
for the functions
$(\Delta_+, \Delta_-)$ 
follows in the same way as in the proof of Theorem~\ref{thm:IntegralEq}.
The time-dependent single-barrier case can be treated in an analogous way.
This completes the proof.
\end{pr}

We will conclude Section~\ref{sec:IntegralEq}
by considering the uniqueness of the solution of the Volterra integral
equation in~(\ref{eq:OneBar})
for a 
time-dependent single barrier 
in the Black-Scholes model.
A much more general result establishing the uniqueness of 
the solution
of the system of Volterra integral equations
of the first kind given in~(\ref{eq:DeltaEq}), 
which requires a detailed analysis of the corresponding 
compact operators,
will be discussed in a subsequent paper.

\begin{prop}
\label{prop:Unique}
Assume that the payoff function
$\phi$
and the barrier
$b:[0,T]\rightarrow\RR_+$
satisfy the hypotheses of Theorem~\ref{thm:IntegralEq}
and let the asset price process
$(S_t)_{t\in[0,T]}$
follow a geometric Brownian motion.
Then integral equation~(\ref{eq:OneBar})
has a unique continuous solution
$\Delta:[0,T]\rightarrow\RR$.
\end{prop}

\begin{pr}
Since equation~(\ref{eq:OneBar})
is linear and, by Theorem~\ref{thm:IntegralEq},
has a continuous solution, it is enough to 
show that the only continuous solution 
$f:[0,T]\rightarrow\RR$
of 
$\int_t^Tq_{u-t}(b(t),b(u))f(u) du=0$
is the obvious one,
i.e.
$f\equiv0$.
In the case of geometric Brownian motion, the integral kernel
$q_{u-t}(b(t),b(u))$
is explicitly given by formula~(\ref{eq:density}).
By Theorem~2.1 
in~\cite{Weiss}
the uniqueness of the solution of the above integral equation
follows, if we prove that the functions
$k(u,t):= \sqrt{u-t} q_{u-t}(b(t),b(u))$
and
$\frac{\partial k}{\partial u}(u,t)$
are continuous
for all 
$u,t\in[0,T]$,
such that 
$u\geq t$,
and that
$k(t,t)$
is non-zero for all
$t\in[0,T]$.

It follows from~(\ref{eq:density})
that the function 
$k(u,t)$
can be expressed as
$$
k(u,t)=\frac{b(u)\sigma}{\sqrt{2\pi}}\exp\left(-\frac{\left(B(u)-B(t)-(\mu-\sigma^2/2)(u-t)\right)^2}{2\sigma^2(u-t)}\right),$$
where
$B(t):=\log b(t)$,
which is clearly continuous for all
$u>t$
and has a non-zero limit, as 
$u$
approaches
$t$,
equal to
$k(t,t)=\frac{b(t)\sigma}{\sqrt{2\pi}}$.
This is a consequence of the Lagrange theorem 
($B(u)-B(t)=B'(\xi_u)(u-t)$
for some 
$\xi_u\in(t,u)$),
applied
to the differentiable function
$B$.
A short calculation shows that the partial derivative
$\frac{\partial k}{\partial u}(u,t)$
exists for all
$u>t$.
The regularity of the function 
$B$
implies that 
$\frac{\partial k}{\partial u}(u,t)$
has a finite limit at
$u=t$
and can therefore be extended to a continuous
function for all
$t,u\in[0,T]$,
such that
$t\leq u$.
This concludes the proof of the proposition.
\end{pr}

%% file: Examples.tex
In this section we will consider some examples that illustrate
the results of 
Theorems~\ref{thm:IntegralEq}
and~\ref{thm:Final}.
We will look at the simplest one in Subsection~\ref{subsec:Simp}.
In~\ref{subsec:SingBar} 
we solve the system of Volterra integral equations~(\ref{eq:DeltaEq})
for the case of constant barriers using Laplace transforms. Subsection~\ref{subsec:Time} 
briefly discusses numerical methods for solving the system~(\ref{eq:DeltaEq})
in the general time-dependent case.

\begin{subsection}{Model-free barrier option price}
\label{subsec:Simp}

It is well-known that a down-and-out
call option struck at
$K$,
with a barrier at the level
$B$,
has a unique model-independent price
if 
$B$
coincides with the strike
$K$
and
if both
the interest rates and dividend yields are zero.
Moreover the barrier option price
equals the price of a forward struck at
$B$,
as is easily seen by the following semi-static replication argument:
since there are no interest rates or dividend yields, 
when the barrier is breached for the first time
the forward contract is worth zero 
and can therefore be sold
at no cost (or gain).
If the barrier is not breached at all,
the two payoffs
clearly coincide.
Since the forward has a model-independent price, the no-arbitrage
principle implies that the barrier option price must equal
$S-B$,
where 
$S$
is the asset price at the current time.
Therefore the delta of the barrier 
option price is identically equal to one
in any model.
In particular the same must be true 
at the barrier.

Let
$C_t(S,K)$
denote the call option price at time
$t$
in the Black-Scholes model.
Equation~(\ref{eq:OneBar})
of Theorem~\ref{thm:IntegralEq}
tells us that the following identity must 
hold
$$B\left(N(\sigma\sqrt{T-t}/2)-N(-\sigma\sqrt{T-t}/2)\right)=\frac{1}{2}\int_t^T
\frac{B\sigma}{\sqrt{2\pi (u-t)}}\exp\left(-\frac{\sigma^2(u-t)}{8}\right)du$$
for all
$t\in[0,T]$,
where the left-hand side equals the Black-Scholes formula for
$C_t(B,B)$
(the function
$N(x)$
is the cumulative normal distribution)
and the integrand on the right-hand side is 
given by~(\ref{eq:density})
and the aforementioned fact
$\Delta\equiv 1$.
Substitution
$x^2=\sigma^2(u-t)/4$
and
a short calculation show that this identity holds for all 
$t\in[0,T]$.
Theorem~\ref{thm:IntegralEq}
therefore implies the following integral representation for the 
linear function
$S\mapsto (S-B)$
$$S-B=C_0(S,B)-\frac{1}{2}\int_0^T
\frac{B\sigma}{\sqrt{2\pi t}}\exp\left(-\frac{\left(\log(B/S)+t\sigma^2/2\right)^2}{2\sigma^2t}\right)dt,$$
where
$C_0(S,B)$
is the Black-Scholes formula.
\end{subsection}

\begin{subsection}{Constant barriers}
\label{subsec:SingBar}
A key distinction between the constant barrier case 
and a time-dependent barrier case, which makes the former
much easier to solve in a semi-analytic form,
is that the kernels of the 
integral operators in equations~(\ref{eq:DeltaEq}) and~(\ref{eq:OneBar})
depend only on the difference of the arguments
$Q(t,u)=Q(u-t)$
when the barriers are constant.
Therefore the delta along the barrier can be obtained
by the following two-step procedure.
First solve an auxiliary integral equation 
where the right-hand side is identically equal to one
using the Laplace
transform method,
which can be applied precisely because the kernel 
depends on the difference of the arguments and the
integral equation is therefore given as a convolution
of two functions.
In the second step 
an explicit
integral representation for the delta along the barrier 
can be constructed
using the solution of the auxiliary equation.
In~\ref{subsec:SingBarConst} we apply this method to single-barrier
options in the Black-Scholes model (see~(\ref{eq:GenSol}) for the explicit formula and~\cite{Integral}, 
Sections~8.4-1 and 8.4-4 for more
details). In~\ref{subsec:DoubBarConst}  we 
generalize this approach to the double-barrier 
case by finding the explicit solution of the system
in~(\ref{eq:DeltaEq}). 

\begin{subsubsection}{Single-barrier options}
\label{subsec:SingBarConst}
Let 
$C_t(S,K)=F_tN(d_+)-KN(d_-)$
denote the discounted value of the European call option in the Balck-Scholes
model,
where the forward is given by
$F_t:=Se^{\mu(T-t)}$,
the drift equals
$$\mu:=r-d\quad
\mathrm{and}\quad
d_\pm:=\log\left(\frac{F_t}{\sigma\sqrt{T-t}}\right)\pm\frac{\sigma\sqrt{T-t}}{2}.$$
Let 
$\alpha:=\frac{(\mu-\sigma^2/2)^2}{2\sigma^2}$
be a positive constant and let the function
$q:\RR_+\rightarrow\RR_+$
equal
$q(t):=\frac{e^{-\alpha t}}{\sqrt{\pi t}}$.
Let
$B$
denote the lower barrier (i.e.
$B<K$)
and let 
$\Delta:[0,T]\rightarrow \RR_+$
be the delta of the option at level 
$B$.
By Theorem~\ref{thm:Final}
we need to solve the integral equation
$C_t(B,K)=\frac{1}{2}\int_t^T\Delta(u) q(u-t)du$
on the interval 
$[0,T]$.
The substitution 
$x:=T-u, y:=T-t$
transforms the equation to
\begin{eqnarray}
\label{eq:LowerBar}
\Psi(y)=\int_0^yf(x) q(y-x)dx,
\>\>\>\>\>\> \>\>\>\>\>\> y\in[0,T],
\end{eqnarray}
where
$\Psi(y):=\frac{2\sqrt{2}}{B\sigma} C_{T-y}(B,K)$
and the unknown function
$f$
is given by
$f(x):=\Delta(T-x).$

As mentioned above, we first solve the auxiliary equation
$1=\int_0^yh(x)q(y-x)dx$.
Recall that the Laplace transform of a function
$h$
is defined by 
$\L(h)(s):=\int_0^\infty e^{-sx}h(x)dx$
for all
$s>0$
such that the integral exists. 
It is obvious that
$\L(1)(s)=\frac{1}{s}$
and a short calculation yields
$\L(q)(s)=\frac{1}{\sqrt{s+\alpha} }$.
By applying the Laplace transform
to both sides of the auxiliary equation
we find 
$\L(h)(s)=\frac{\sqrt{s+\alpha}}{s}$,
since 
the right-hand side 
equals
$\L(h*q)(s)=\L(h)(s)\L(q)(s)$
by the famous property of the Laplace transform.
The function
$(h*q)(y):=\int_0^yh(x)q(y-x)dx$
in this formula
denotes the convolution	of 
$h$
and
$q$.
Note that both equation~(\ref{eq:LowerBar}) and the auxiliary 
equation can be represented in the following way:
$\Psi(y) = (f*q)(y)$
and
$1 = (h*q)(y)$.
This simple observation will be useful in Subsection~\ref{subsec:DoubBarConst}.

The task now is to compute the inverse Laplace transform
$\L^{-1}$,
which is defined as an integral along a path in the complex
plane,
of the function
$s\mapsto\frac{\sqrt{s+\alpha}}{s}$.
Instead of using the definition of
$\L^{-1}$
we observe the following elementary identities
$$ \frac{1}{\sqrt{s+\alpha}} +\sqrt{\alpha} \frac{\sqrt{\alpha}}{s\sqrt{s+\alpha}}=
\frac{\sqrt{s+\alpha}}{s}\>\>\>\>\>\>\>\>\>\>\>\>\>\>\>\> \mathrm{and}\>\>\>\>\>\>\>\>\>\>\>\>\>\>\>\>    
\L(x\mapsto E(\sqrt{\alpha x}))(s)= \frac{\sqrt{\alpha}}{s\sqrt{s+\alpha}},$$
where
$E(x):=\frac{2}{\sqrt{\pi}}\int_0^xe^{-v^2}dv$
is the error function.
The first identity is obvious and
the second follows from the discussion above
upon noticing that the function
$x\mapsto E(\sqrt{\alpha x})$ 
can be expressed as a convolution
$E(\sqrt{\alpha x})=(\sqrt{\alpha}*q)(x)$.
By applying the inverse Laplace transform to the first identity
it now follows that the solution of the auxiliary equation
is
$h(x)=q(x)+\sqrt{\alpha} E(\sqrt{\alpha x})$. 
Fubini's theorem and the auxiliary equation can now be used to
verify that the function
\begin{eqnarray}
\label{eq:GenSol}
f(x):= h(x)\Psi(0)+ (h*\Psi')(y)
\end{eqnarray}
solves integral equation~(\ref{eq:LowerBar}).
Since
$\Psi(0)=0$
and we have the formula 
$$\Psi'(y)=\frac{2\sqrt{2}}{\sigma} e^{\mu y}\left(\mu N(d_+)+\frac{\sigma}{2\sqrt{y}}N'(d_-)\right),$$
the delta along the barrier
$\Delta:[0,T]\rightarrow \RR_+$
can in the case of a down-and-out call be expressed as a convolution of two explicit functions
$\Delta(t)= (h*\Psi')(T-t).$
It is well-known that in the symmetric case when
$\mu = 0$,
the down-and-out barrier option price in the Black-Scholes model is given
by 
$C_0(S,K)-\frac{K}{B}P_0(S,\frac{B^2}{K})$,
where 
$P_0(S,\frac{B^2}{K})$
is a put option struck at 
$\frac{B^2}{K}$
(see~\cite{Lipton}, page~454, equation~(12.3)).
Theorem~\ref{thm:Final}
therefore yields the equation
$P_0(S,\frac{B^2}{K})= e^{-r(T-t)}\frac{B}{K} \int_0^T\Delta(t)q_t(S,B)dt,$
where
$q_t(S,B)$
is given in~(\ref{eq:density}).

The key observation is that the procedure described
here works for an up-and-out call option in precisely the same way.
The auxiliary equation again takes the form 
$1=(h*q)(y)$
and hence has the same solution as before. 
Function 
$f$
defined in~(\ref{eq:GenSol})
solves integral equation~(\ref{eq:LowerBar}),
where the function 
$\Psi$
is redefined appropriately. In the case of an
up-and-out call we have
$\Psi(y)=-\frac{2\sqrt{2}}{B\sigma}\EE_{T-y,B}[(S_T-K)^+I_{\{S_T\leq B\}}]$
where the barrier 
$B$
is larger than the strike
$K$.
In fact the same procedure works for the class 
of linear diffusions considered in Theorem~\ref{thm:Final},
as long as one is prepared to calculate (numerically or otherwise)
the inverse Laplace transform
of the function 
$s\mapsto \L(q)(s)/s$.
The function
$q:[0,T]\rightarrow \RR_+$
would in this case depend on the underlying diffusion through
formula~(\ref{eq:KernelFun}) with $x$ and
$y$ equal to the barrier level $B$.
\end{subsubsection}

\begin{subsubsection}{Double-barrier options}
\label{subsec:DoubBarConst}
Let
$B_-$
and
$B_+$
denote the lower and upper barrier respectively
and let functions
$\Psi_1, \Psi_2:[0,T]\rightarrow\RR$
be given by
$\Psi_1(y):=2\varphi(T-y,B_+)$
and
$\Psi_2(y):=2\varphi(T-y,B_-)$,
where
$\varphi$
represents the discounted value of the European payoff
(see
Theorem~\ref{thm:IntegralEq}).
By introducing the change of variable
$x:=T-u$
as in the previous subsection
and denoting
$f_1(x):=\Delta_+(T-x)$,
$f_2(x):=\Delta_-(T-x)$,
we can express equation~(\ref{eq:DeltaEq})
using the linear operator
$\K:L^1([0,T])\times L^1([0,T]) \rightarrow L^1([0,T])\times L^1([0,T])$
in the following way
\begin{eqnarray}
\label{eq:MatrixEq}
\begin{pmatrix}
\Psi_1 \\
\Psi_2
\end{pmatrix}
=
\K
\begin{pmatrix}
f_1 \\
f_2
\end{pmatrix}\!,
\>\>\>\>
\mathrm{where}
\>\>\>\>
\K
\begin{pmatrix}
f_1 \\
f_2
\end{pmatrix}(y):=
\begin{pmatrix}
\int_0^yQ_{11}(y-x)f_1(x)dx +  \int_0^yQ_{12}(y-x)f_2(x)dx \\
\int_0^yQ_{21}(y-x)f_1(x)dx +  \int_0^yQ_{22}(y-x)f_2(x)dx \\
\end{pmatrix}\!.
\end{eqnarray}
The functions
$Q_{ij}:[0,T]\rightarrow \RR$,
$i,j\in\{1,2\}$,
are the coordinates of the matrix in~(\ref{eq:matrix})
and can be expressed as functions
of one variable precisely because the
barriers are constant in time.

Recall that convolution
can be used to make the Banach space
$L^1([0,T])$
into a commutative Banach algebra,
since the function
$(u*v)(y)=\int_0^yu(y-x)v(x)dx$
is an element of
$L^1([0,T])$
for any 
$u,v\in L^1([0,T])$.
Using this multiplicative structure 
and the definition in~(\ref{eq:MatrixEq})
we can express the linear 
operator 
$\K$
in the following way
$$
\K
\begin{pmatrix}
f_1 \\
f_2
\end{pmatrix} =
\begin{pmatrix}
Q_{11} & Q_{12} \\
Q_{21} & Q_{22} \\
\end{pmatrix}*
\begin{pmatrix}
f_1 \\
f_2
\end{pmatrix}.
$$

The integral equation in~(\ref{eq:MatrixEq})
can now be solved in two steps.
The first step consists of finding 
the functions
$h_{ij}:[0,T]\rightarrow\RR$,
$i,j\in\{1,2\}$,
which satisfy the identity
\begin{eqnarray}
\label{eq:AuxEq}
\begin{pmatrix}
1 & 0 \\
0 & 1 \\
\end{pmatrix}
=
\begin{pmatrix}
Q_{11} & Q_{12} \\
Q_{21} & Q_{22} \\
\end{pmatrix}*
\begin{pmatrix}
h_{11} & h_{12} \\
h_{21} & h_{22} \\
\end{pmatrix}.
\end{eqnarray}
Note that the product of any pair of coordinate functions in this expression
is given by 
their convolution.
Note also that the solution of this auxiliary equation
depends solely on the barrier levels
$B_+, B_-$
and is independent of
the payoff of the option we are trying to 
price.
By applying the Laplace transform 
$\L$
to each coordinate of this equation,
we obtain a linear system for the functions
$\L(h_{ij})$,
where multiplication is defined using a point-wise
product rule:
$$
\begin{pmatrix}
\frac{1}{s} & 0 \\
0 &  \frac{1}{s} \\
\end{pmatrix}
=
\begin{pmatrix}
\L(Q_{11})(s) & \L(Q_{12})(s)  \\
\L(Q_{21})(s)  & \L(Q_{22})(s)  \\
\end{pmatrix}
\begin{pmatrix}
\L(h_{11})(s)  & \L(h_{12})(s)  \\
\L(h_{21})(s)  & \L(h_{22})(s)  \\
\end{pmatrix}.
$$
Assuming that the determinant 
$(\L(Q_{11}) \L(Q_{22})-\L(Q_{12})\L(Q_{21}))(s)$
is non-zero for all 
$s>0$,
we can explicitly solve this system of equations.
In order to obtain the functions
$h_{ij}:[0,T]\rightarrow\RR$,
$i,j\in\{1,2\}$,
we need to perform Laplace inversion on each of the four
coordinates of the solution of the linear system.

Once the auxiliary equation in~(\ref{eq:AuxEq})
has been solved, we can express the solution of the 
original integral equation in~(\ref{eq:MatrixEq})
in the following way
$$
\begin{pmatrix}
f_1(x) \\
f_2(x)
\end{pmatrix}=
\begin{pmatrix}
h_{11}(x) & h_{12}(x) \\
h_{21}(x) & h_{22}(x) \\
\end{pmatrix}
\begin{pmatrix}
\Psi_1(0) \\
\Psi_2(0)
\end{pmatrix}+
\begin{pmatrix}
h_{11} & h_{12} \\
h_{21} & h_{22} \\
\end{pmatrix}*
\begin{pmatrix}
\Psi_1' \\
\Psi_2'
\end{pmatrix}(x).
$$
Since in our case we have 
$\Psi_1(0)=\Psi_2(0)=0$,
the deltas at the upper 
and lower 
barriers
are given by the formulae
$\Delta_+(t)=(h_{11}*\Psi_1')(T-t) + (h_{12}*\Psi_2')(T-t)$
and
$\Delta_-(t)=(h_{21}*\Psi_1')(T-t) + (h_{22}*\Psi_2')(T-t)$
respectively.
Representation~(\ref{eq:PriceRepNonCont})
of the double-barrier option price in Theorem~\ref{thm:Final}
can now be applied.
\end{subsubsection}
\end{subsection}

\begin{subsection}{Time-dependent barrier options}
\label{subsec:Time}
In case of general time-dependent barriers not much can be said analytically
about the structure of the solutions of the system of integral
equations in~(\ref{eq:DeltaEq}). 
However the trapezoidal product integration method, described in~\cite{Weiss},
can be applied directly to the single-barrier problem. The substitutions 
$y:=T-t$
and
$x:=T-u$,
used in Subsection~\ref{subsec:SingBar},
transform equation~(\ref{eq:OneBar}) into a generalised Abel equation 
with a weakly singular kernel 
$$k(y,x):=\mp\frac{\sqrt{y-x}}{2}q_{y-x}(b_\pm(T-y),b_\pm(T-x)),$$
where the function
$b_\pm:[0,T]\rightarrow \RR_+$
is either a lower or an upper barrier
and the function
$q$
is given in~(\ref{eq:KernelFun}).
Using the notation
$\Psi(y):=\varphi(T-y,b_\pm(T-y))$
for the discounted value of the European payoff
(see Theorem~\ref{thm:IntegralEq}
for the precise definition of
$\varphi$)
and
$f(x):=\Delta_\pm(T-x)$
for the unknown function in our integral equation,
we can rewrite~(\ref{eq:OneBar}) 
as follows:
\begin{eqnarray}
\label{eq:IntBarOne}
\Psi(y)=\int_0^y\frac{k(y,x)}{\sqrt{y-x}}f(x)dx.
\end{eqnarray}
It follows from the representation 
of the density of a linear diffusion given 
in~\cite{FriedmanPar}, Section~1.2, equation~(2.8),
that 
$k(y,y):=\lim_{x\nearrow y}k(y,x)$
exists and is non-zero since we are assuming that the barrier function
is differentiable. This statement is clear for 
geometric Brownian motion,
when the function
$q$
is
given by~(\ref{eq:density}),
and essentially the same proof can be used for a general
linear diffusion once we apply the representation given in~\cite{FriedmanPar}.
The observation 
that 
$0<k(y,y)<\infty$,
for all
$y\in[0,T]$,
is of utmost importance because it makes the lower-triangular system of linear equations
in~\cite{Weiss}
non-singular.

The main theorem of~\cite{Weiss}
says that the solution of the lower-triangular linear system, given by equation~(4.1) 
on page~179 of the same paper,
converges to the solution of the integral equation~(\ref{eq:IntBarOne}) at the order of
$O(h^2)$,
where
$h$ 
is the distance between the consecutive points in the discretization of 
$[0,T]$.
This convergence result
assumes some
regularity properties
of the solution 
$f$,
such as continuity 
on the entire interval
$[0,T]$,
which are in general not satisfied in our context.
The situation is improved if we work in the domain 
of Theorem~\ref{thm:IntegralEq}.
In other words when faced with a discontinuous payoff function
$\phi$,
we can first approximate 
it 
by a smooth function
$\phi_n$,
as in the proof of Theorem~\ref{thm:Final},
and then solve the linear system from~\cite{Weiss}
which corresponds to the derivative that delivers
$\phi_n$.
This procedure introduces an additional numerical error
since we are pricing the ``wrong'' derivative, but improves 
the convergence speed of the solution of the lower-triangular
linear system 
from~\cite{Weiss}.
It follows from the construction 
of
$\phi_n$
in the proof of Theorem~\ref{thm:Final}
that the price of the 
barrier option with the payoff
$\phi_n$
converges uniformly in
$(t,S_t)$
to the price of the same barrier option with 
the payoff
$\phi$.
The stability of the proposed numerical algorithm
will be the subject of future research.

The double-barrier case can be dealt with similarly 
upon noticing that the functions
$q_{u-t}(b_+(t),b_-(u))$
and
$q_{u-t}(b_-(t),b_+(u))$,
which appear ``off the diagonal''
in the kernel of
the system of Volterra equations given in~(\ref{eq:matrix}),
are smooth and bounded for all
$t,u\in[0,T]$
such that
$t\leq u$.
In other words we 
can extend the 
$n$-dimensional lower-triangular system from \cite{Weiss},
used to solve integral equation~(\ref{eq:IntBarOne}),
to a 
$2n$-dimensional linear system by representing 
the integrals against functions
$q_{u-t}(b_+(t),b_-(u))$
and
$q_{u-t}(b_-(t),b_+(u))$
using the standard trapezoidal method
(which can be expressed as
matrix vector multiplication).
By expressing the solution vector
in the following way
$(\Delta_+(t_1),\Delta_-(t_1),\ldots,\Delta_+(t_n),\Delta_-(t_n))^T$,
where
$(t_i)_{i=1,\ldots,n}$
is an
increasing sequence such that
$t_1=0$
and
$t_n=T$,
the 
$2n$-dimensional linear system we need to solve
becomes lower-triangular because of the identities
$\lim_{u\searrow t}q_{u-t}(b_+(t),b_-(u))= \lim_{u\searrow t}q_{u-t}(b_-(t),b_+(u))=0$.
There are a number of algorithms designed to solve this kind of linear system very 
quickly and accurately
(see for example~\cite{TriangSys}). 
Their implementations usually rely on numerical libraries like
BLAS and LAPACK for the calculations. These numerical libraries 
are highly optimised and 
can be called directly from C++
(see~\cite{LAPACK}
for more information on LAPACK, also
available at
http://www.netlib.org/lapack/lug/).
However 
the question of 
implementation and the optimal choice 
of algorithm 
for solving our lower-triangular linear systems
requires further numerical investigation.
\end{subsection}

%% file: Conclusion.tex
In this paper we have obtained an integral representation of 
the difference between the time-dependent
double-barrier
option price and the price of a European option
with the same payoff. Theorems~\ref{thm:IntegralEq} and~\ref{thm:Final}
give the precise formulae in terms of the double-barrier option deltas
$(\Delta_+,\Delta_-):[0,T]\rightarrow\RR\times\RR$
at the barriers 
(see (a) in Lemma~\ref{lem:regularD}) for the precise definitions of these functions),
which solve the system of Volterra integral equations of the first
kind
in~(\ref{eq:DeltaEq}).
It follows by construction that the system of integral equations
in~(\ref{eq:DeltaEq})
has a solution. The most natural question is the one of uniqueness of
solution, which is equivalent to the question of whether 
zero
is an eigenvalue of the compact linear integral operator with 
a weakly singular kernel given in~(\ref{eq:matrix}). 
In general, compact operators exhibit both kinds of behaviour and 
the standard technique of transforming an integral equation of the first kind
$Kf=\Psi$
to an integral equation of the second kind 
$f-Lf=\Psi$,
where
$K$
and
$L$
are related integral operators
(see~\cite{Integral}, Section~8.3),
does not apply in our setting because of the weak singularity of our kernel.
The transformation of the problem is desirable because 
we can use the Fredholm alternative to analyse 
the kernel
of the operator
$I-L$,
where 
$L$
is compact
and 
$I$
is the identity operator.
In spite of this difficulty this general approach can be made to work 
in the case of equation~(\ref{eq:DeltaEq}) by careful inspection
of the construction of the functions in the integral kernel and
the uniqueness of the solution can be proved, at least for payoff functions
that satisfy the regularity conditions of Lemma~\ref{lem:regularD}. 
The proof will be given in a subsequent publication.

If we assume that the Volterra integral equation of the first 
kind in~(\ref{eq:DeltaEq}) has a unique solution, then 
representation~(\ref{eq:PriceRep}) for the time-dependent double-barrier option price from 
Theorem~\ref{thm:IntegralEq}
implies that the value of this path-dependent derivative 
depends only on the one-dimensional distributions in the risk-neutral 
measure of the underlying process
$S_t$.
It is well-known that having the vanilla option prices for all strikes and
all maturities is equivalent to having the one-dimensional distributions of
$S_t$.
Theorem~\ref{thm:IntegralEq}
therefore provides an explicit link between the vanilla option prices and the barrier
option prices for all reasonably smooth barriers and a wide class of local
volatility models.
The fact that
vanilla option prices determine uniquely the barrier
option prices in the 
world of local volatility models has been known since the
seminal work of Bruno Dupire~\cite{Dupire} where, under certain
regularity conditions, 
a PDE for 
the local volatility function
$x\mapsto \sigma(x)$
is derived from the vanilla option prices.
This in turn determines the risk-neutral dynamics of
$S_t$
and therefore the prices of all path-dependent derivatives but does not
yield an explicit relationship.

In this paper we have discussed a barrier pricing problem 
without rebates. In general a barrier can pay a
contract defined 
rebate
$F_+(t)$
(resp. 
$F_-(t)$)
if at time
$t\in(0,T)$
the asset price process
$S_t$
equals the barrier level
$b_+(t)$
(resp.
$b_-(t)$)
for the first time since inception.
It is not difficult to see that under 
some additional technical assumptions 
on the functions 
$F_\pm:[0,T]\rightarrow\RR$,
the change-of-variable formula from~\cite{Peskir}
can be applied and a similar technique to the one
used to prove Theorem~\ref{thm:IntegralEq}
yields an integral representation
of the time-dependent double-barrier option price with rebate.
In fact the final formula is very similar to the
one in~(\ref{eq:PriceRep}), 
with 
$\varphi(0,S_0)$
replaced by the sum of the expectation of the European 
payoff and certain integrals over the time-interval
$[0,T]$
of the rebate functions
$F_\pm$.

We have seen that purely 
probabilistic concepts such as local time
and the generalized It\^{o} formula proved by
Peskir in~\cite{Peskir}	
can be used to obtain a new structure
in the barrier option pricing problem,
which can then be applied to the pricing and hedging
of double-barrier options in local
volatility models. This structure consists of
two deterministic functions
$\Delta_\pm:[0,T]\rightarrow\RR$,
which represent the deltas at the two
barriers.
It is intuitively clear that the same structure
exists in stochastic volatility models.
Since Peskir's formula has been generalized
to higher dimensions
in~\cite{Peskir_MultDim} 
for all (possibly discontinuous) semimartingales,
a generalisation of the approach presented 
here might be feasible. The multidimensional 
change-of-variable
formula
in~\cite{Peskir_MultDim}
is both surprising and satisfactory,
not just because it applies to all semimartingales
but because, under natural deterministic conditions
on the value function, the resulting formula is a direct
extension of the one-dimensional formula in~\cite{Peskir}.
However a direct application of the 
formula
in~\cite{Peskir_MultDim}
is not possible in our case, because it requires the existence
of 
a regular extension of the value function across the
boundaries of its natural domain, a question that requires
some investigation. 
The issue of which quantity
one could represent in 
terms of deltas at the barriers in the higher dimensional 
case (the value 
function itself cannot be represented) provides in our view an additional
interesting problem 
for future research.

%
%
%
%

%% file: changeofvariable.tex
In this section we establish a mild generalisation
of the change-of-variable formula given in Theorem 3.1
of~\cite{Peskir}.
In fact Theorem~\ref{thm:Goran}
is implicitly proved in~\cite{Peskir}.
Since Theorem~\ref{thm:Goran}
is central to our analysis,
for complicity we give a proof based on a direct application of
Theorem 3.1
and
Remark 2.5
in~\cite{Peskir}.

Let 
$X:=(X_t)_{t\in[0,T]}$
be an It\^{o}
diffusion that solves the following stochastic differential
equation
$dX_t=M(X_t) dt+\Sigma(X_t)dW_t$,
where
$M(x):=\mu x$
and
$\Sigma(x):=x\sigma(x)$,
and
let
$b_\pm:[0,T]\rightarrow\RR$
be two continuous functions of finite variation
satisfying
$b_-(t)<b_+(t)$
for all 
$t\in[0,T]$.
As before we set 
$C:=\{(t,x)\in[0,T)\times\RR;\>b_-(t)<x<b_+(t)\}$,
$B_+:=\{(t,x)\in[0,T)\times\RR;\>x>b_+(t)\}$
and
$B_-:=\{(t,x)\in[0,T)\times\RR;\>x<b_-(t)\}$.
Let
$F:[0,T]\times\RR\rightarrow\RR$
be a continuous function which is
$C^{1,2}$
on the open subset
$B_-\cup C\cup B_+$
of
$[0,T]\times\RR$.
Given a function 
$g:[0,T]\rightarrow \RR$
of bounded variation let 
$V(g)(t)$
denote the total variation of
$g$
on
$[0,t]$
for any 
$t\leq T$.

\begin{thm}
\label{thm:Goran}
Let
$\Sigma(x)>0$
for all 
$x\in(0,\infty)$
such that
$(t,x)\in \overline{C}$
and assume that 
$F_t+M F_x+\frac{\Sigma^2}{2}F_{xx}$
is locally bounded on
$B_-\cup C\cup B_+$,
the limit
$F_x(s,b_\pm(s)\pm):=\lim_{\epsilon\searrow0}F_x(s,b_\pm(s)\pm\epsilon)$
is uniform in
$s\in[0,t]$
and that 
$\sup_{0<\epsilon<\delta}V(F(\cdot,b_\pm(\cdot)\pm\epsilon))(t)<\infty$
for some 
$\delta>0$
and any combination of sings 
$+$
and 
$-$.
Then the following change-of-variable formula holds:
\begin{eqnarray*}
F(t,X_t) & = & F(0,X_0) + \int_0^t(F_t+M F_x+\frac{\Sigma^2}{2}F_{xx})(s,X_s)
               I_{\{X_s\neq b_-(s),X_s\neq b_+(s)\}}ds \\
    & &      +\int_0^t(\Sigma F_x)(s,X_s)I_{\{X_s\neq b_-(s),X_s\neq b_+(s)\}}dW_s
                  \\
         &   & + \frac{1}{2}\int_0^t(F_x(s,X_s+)-F_x(s,X_s-))I_{\{X_s=b_-(s)\}}dL_s^{b_-}(X)\\
         &   & + \frac{1}{2}\int_0^t(F_x(s,X_s+)-F_x(s,X_s-))I_{\{X_s=b_+(s)\}}dL_s^{b_+}(X).
\end{eqnarray*}
\end{thm}

For 
the definition of 
the local time of
$X$
at the curve 
$b$,
$L_t^b(X)$,
see page~\pageref{p:localtime}.

\begin{pr}
We start by defining 
continuous functions
$F^\pm:[0,T]\times\RR\rightarrow\RR$
which satisfy the hypothesis of Theorem 3.1 in~\cite{Peskir}
and then apply the theorem to obtain the formula above.
Since functions
$b_\pm$
are continuous,
the images
$b_\pm([0,T])$
are disjoint compact subsets in
$\RR^2$
with strictly positive distance.
Hence
there exist
$\epsilon>0$
such that 
$4\epsilon<b_+(t)-b_-(t)$
for all
$t\in[0,T]$.
It is clear that there exist 
smooth functions
$c_\pm:[0,T]\rightarrow \RR$
that satisfy 
$b_-(t)<c_-(t)<b_-(t)+\epsilon$
and
$b_+(t)-\epsilon<c_+(t)<b_+(t)$
for all 
$t\in[0,T]$.
In particular it follows that
$c_+(t)-c_-(t)>2\epsilon$.

We now define continuous functions
$F^+$
and 
$F^-$,
which are
$C^{1,2}$
everywhere in 
$[0,T]\times\RR$
except along the curves
$b_+$
and
$b_-$
respectivly,
by the formulae
$$F^+(t,x):= \left\{ \begin{array}{ll}
F(t,x) & \textrm{if $x\geq c_-(t)$,}\\
f^+(t,x) & \textrm{if $x< c_-(t)$,}
\end{array} \right.\>\>\>\>\>\>\>\>
F^-(t,x):= \left\{ \begin{array}{ll}
F(t,x) & \textrm{if $x\leq c_+(t)$,}\\
f^-(t,x) & \textrm{if $x> c_+(t)$.}
\end{array} \right.$$
The function
$f^+$
(resp.
$f^-$)
is a 
$C^{1,2}$
extension of 
$F$
across the smooth boundary
$c_-$
(resp. 
$c_+$),
which exists because 
$F$
is 
$C^{1,2}$
by assumption 
on
the domain
$C$.
Note also that the functions
$f^\pm$ 
are non-unique. Given these definitions
of 
$F^\pm$,
the only discontinuities of the derivatives
are the ones inherited from the original
function
$F$
along the curves
$b_\pm$
respectively.
Since 
$F$
satisfies the conditions in Theorem~\ref{thm:Goran},
the functions
$F^\pm$
also satisfy the assumptions of Theorem 3.1 in~\cite{Peskir},
which therefore implies the following formulae for any fixed time
$t\in[0,T]$:
\begin{eqnarray}
F^\pm(t,X_t) & = & F^\pm(0,X_0) + \int_0^t(F^\pm_t+M F^\pm_x+\frac{\Sigma^2}{2}F^\pm_{xx})(s,X_s)
           I_{\{X_s\neq b_\pm(s)\}}ds+\int_0^t(\Sigma F^\pm_x)(s,X_s)
              I_{\{X_s\neq b_\pm(s)\}}dW_s \nonumber
                  \\
         &   & 
             + \frac{1}{2}\int_0^t\left(F^\pm_x(s,X_s+)-F^\pm_x(s,X_s-)\right)I_{\{X_s=b_\pm(s)\}}dL_s^{b_\pm}(X),
             \label{eq:LocTimeFormula}
\end{eqnarray}
where the signs 
$\pm$
are simultaneously equal to either 
$+$
or
$-$.

Let
$(\F_t)_{t\in[0,T]}$
denote the filtration
of the Brownian motion
$(W_t)_{t\in[0,T]}$
that satisfies the usual conditions. 
Since the processes on both sides 
of the equality in~(\ref{eq:LocTimeFormula})
have continuous paths we can assume that they are 
indistinguishable and therefore substitute 
fixed time 
$t\in[0,T]$
with any stopping time relative to
$(\F_t)_{t\in[0,T]}$.

We now define an increasing sequence of
stopping times in the following way:
$\rho_1:=t\wedge\inf\{s;X_s=c_+(s)\},\>\>
\rho_2:=t\wedge\inf\{s>\rho_1;X_s=c_-(s)\}$
and
$\rho_{2n+1}:=t\wedge\inf\{s>\rho_{2n};X_s=c_+(s)\},\>\>
\rho_{2n+2}:=t\wedge\inf\{s>\rho_{2n+1};X_s=c_-(s)\}$
(we are assuming wlog that 
$X_0<c_+(0)$).
Note that for any
$s$
in
$[\rho_{2n+1},\rho_{2n+2}]$
(resp. in
$[\rho_{2n},\rho_{2n+1}]$)
the value of the random variable 
$X_s$
is strictly above 
$b_-(s)$
(resp. below
$b_+(s)$)
and therefore
$F(s,X_s)$
equals
$F^+(s,X_s)$
(resp.
$F^-(s,X_s)$).
We also have
$\lim_{n\rightarrow\infty}\rho_n=t$
a.s.
and,
for almost all paths of
$X$,
$\rho_n=t$
for some 
$n\in\NN$
(this follows from the inequality 
$c_+(s)-c_-(s)>2\epsilon$,
for all
$s\in[0,T]$,
and the fact that the expectation
of the upcrossing number of our
semimartingale is finite, cf.~\cite{Karatzas},
Theorem 1.3.8 (iii)).
For a fixed 
$t\in[0,T]$
we have a telescoping representation
\begin{eqnarray*}
F(t,X_t)-F(0,X_0) & = & \sum_{n=0}^\infty\left(F(\rho_{2n+2},X_{\rho_{2n+2}})
-F(\rho_{2n+1},X_{\rho_{2n+1}})+F(\rho_{2n+1},X_{\rho_{2n+1}})-F(\rho_{2n},X_{\rho_{2n}})
 \right) \\
 & = & \sum_{n=0}^\infty (F^+(\rho_{2n+2},X_{\rho_{2n+2}})
-F^+(\rho_{2n+1},X_{\rho_{2n+1}}))  +
\sum_{n=0}^\infty (F^-(\rho_{2n+1},X_{\rho_{2n+1}})
-F^-(\rho_{2n},X_{\rho_{2n}})),
\end{eqnarray*}
where
$\rho_0:=0$.
We are allowed to reshuffle the summands in this path-wise identity since
for almost all paths the sums consist of finitely many summands.
The theorem now follows by applying formula~(\ref{eq:LocTimeFormula})
between the 
stopping times 
for 
$(\F_t)$
to the summands in the last expression and collecting the terms. 
\end{pr}

%% file: analyticity.tex
Let
$\phi:\RR_+\rightarrow\RR_+$
be a payoff function
that  is continuous on a complement of a finite set
where it is right-continuous with 
left limits. In particular 
$\phi$
is continuous at zero.
An important example is
$\phi(x)=(x-K)^+I_{(B_-,B_+)}(x)$
for some constants
$B_-<K<B_+$
and 
$x\in\RR_+$.
The diffusion
$X=(X_t)_{t\in[0,T]}$,
specified by the time-homogeneous SDE 
$dX_t=M(X_t) dt+\Sigma(X_t)dW_t$
with linear drift
$M(x):=\mu x$
and a locally Lipschitz diffusion coefficient
$\Sigma(x):=x\sigma(x)>0$, 
for
$x\in(0,\infty)$,
is as described in the beginning of Section~\ref{sec:IntegralEq}.
Using the notation from
appendix~\ref{sec:Goran}
we define
a family of  
stopping times 
$\tau_t$,
for any 
$t\in[0,T]$,
by
$\tau_t:=\inf\{v\in[0,T-t];\>\>X_{t+v}\in\RR-(b_-(t+v),b_+(t+v))\}$,
where
the boundary functions
$b_\pm:[0,T]\rightarrow\RR$
are continuous and twice-differentiable
in the interval
$(0,T)$.
We also consider the case where either 
$b_+$
or
$b_-$
are not present
to capture the single-barrier case.
If
there is no upper barrier
we assume in addition that
$\phi(X_T)\in L^1(\Omega,\QQ)$.
The discounted barrier price process is a 
martingale given by
$V_t:=\EE\left[\phi(X_{\tau_0})I_{\{\tau_0=T\}}\arrowvert \F_t\right]$
where the filtration
$(\F_t)_{t\in[0,T]}$
is as described in the beginning of 
Section~\ref{sec:IntegralEq}.
The identity
$I_{\{\tau_0=T\}}=I_{\{\tau_0>t\}}I_{\{\tau_t=T-t\}}$,
the facts 
$\{\tau_0>t\}\in\F_t$,
$\tau_0=t+\tau_t$
on the set
$\{\tau_0>t\}$
and the Markov property of
$X$
imply the following path-wise representation for the barrier
price
$V_t=I_{\{\tau_0>t\}}Z(t,X_t)$,
where the function 
$Z:[0,T]\times\RR\rightarrow\RR$
is given by
\begin{eqnarray}
\label{eqn:BarrierPrice}
Z(t,x):=\EE_{t,x}\left[\phi(X_{t+\tau_t})I_{\{t+\tau_t=T\}}\right].
\end{eqnarray}
Here the process 
$X$
starts at time
$t$
with value
$X_t=x$.

It is often stated that the barrier option price satisfies a certain
PDE with absorbing boundary conditions. Such statements are in fact 
referring to the analyticity properties of 
the function
$Z$,
which we make precise and prove in Theorem~\ref{thm:analyticity}.

\begin{thm}
\label{thm:analyticity}
Let 
$\L g(t,x):=\left(g_t+
M g_x+
\frac{\Sigma^2}{2}g_{xx}\right)(t,x)$
be the infinitesimal generator of the diffusion
$Y=(Y_t)_{t\in[0,T]}$,
where
$Y_t:=(t,X_t)$
and
the process 
$X$
is as described above.
Let the set 
$C$
be as defined in appendix~\ref{sec:Goran}
and
assume that
$\overline{C}$
(the closure is taken in the space
$\RR\times[0,T]$)
is contained in
$(0,\infty)\times[0,T]$.
Then, under the above hypothesis
on the payoff
$\phi$
and
barriers
$b_\pm$,
the function 
$Z$
given by~(\ref{eqn:BarrierPrice})
is continuous on the set
$\overline{C}-(\RR\times\{T\})$
and solves the following parabolic boundary value problem:
\begin{eqnarray*}
\L Z(t,x) & = & 0\>\>\>\mathit{for}\>\>\> (t,x)\in C,\\
Z(T,x) & = & \phi(x)\>\>\>\mathit{for}\>\>\> x\in (b_-(T),b_+(T)),\\
Z(t,b_\pm(t)) & = & 0 \>\>\>\mathit{for}\>\>\> t\in [0,T].
\end{eqnarray*}
In the single-barrier case, the local behaviour and the terminal condition
satisfied by the function 
$Z$
remain the same, but the boundary conditions change as follows:
for an up-and-out option the boundary conditions
are 
$Z(t,b_+(t))  =  0$
and
$Z(t,0)  =  \phi(0)$,
$t\in [0,T]$,
and in a 
down-and-out case we have
$Z(t,b_-(t))  =  0$
for all
$t\in [0,T]$.
\end{thm}          

\begin{pr} 
Assume first that both barriers are present.
If in addition we assume that 
$\phi:[b_-(T),b_+(T)]\rightarrow\RR$
is continuous and satisfies
$\phi(b_-(T))=\phi(b_+(T))=0$,
then by Theorems 6.3.6 and 6.5.2 
in~\cite{Friedman},
function
$Z$
defined by~(\ref{eqn:BarrierPrice})
is the solution of the parabolic PDE
and satisfies the required boundary conditions.
In particular 
$Z$
is continuous on 
$\overline{C}$.

Assume now that
the payoff 
$\phi$
is discontinuous. 
By assumption 
$\phi$
only has
a finite number of bounded jumps. Hence we can 
express it as 
$\phi=\lim_{n\rightarrow\infty}\phi_n$,
where the functions 
$\phi_n$
are continuous with
$\phi_n(b_-(T))=\phi_n(b_+(T))=0$
for all 
$n\in\NN$
and the convergence is uniform on the 
complement of any neighbourhood 
of the 
discontinuities of 
$\phi$.
In fact we can choose the functions
$\phi_n$
so that there exists a decreasing 
sequence of open sets
$N_n\subset\RR$,
such that the intersection
$\cap_{n=1}^\infty N_n$
equals the set of discontinuities of
$\phi$,
and
$\phi(x)=\phi_n(x)$
on the complement
of
$N_n$
for all
$n\in\NN$.
In the obvious notation we get
$\arrowvert Z(t,x)-Z_n(t,x)\arrowvert\leq
A\EE_{t,x}[I_{N_n}(X_T)]$,
where
$A$
is some constant independent 
of
$n$
which exists since
$\phi$
is a bounded function.
As usual
$I_{N_n}$
denotes the indicator function of the set
$N_n$.
Since the random variable
$X_T$
has a density in the set 
$(0,\infty)$
which is smooth in the parameter
$(t,x)$
(see the discussion preceding Theorem~\ref{thm:IntegralEq}
in Section~\ref{sec:IntegralEq}),
the right-hand side of the last inequality
goes to zero uniformly on some neighbourhood of the point
$(t, x)$.
This implies that 
$Z$
is a limit of a uniformly convergent sequence
of continuous functions and is therefore continuous
on the complement of the finite set of
discontinuities of the payoff
$\phi$.
Note that
$Z(t,b_\pm(t)) =\lim_{n\rightarrow\infty}Z_n(t,b_\pm(t))=0$
for all
$t\in[0,T)$
and that 
$Z(T,x)=\phi(x)$
by definition.

We now need to prove that 
$Z$
is in
$C^{1,2}(C)$
and that it satisfies the PDE
$\L Z = 0$,
where 
$\L$
is the infinitesimal generator of the diffusion 
$Y$.
These are local properties of the function
$Z$
and it is therefore enough to show that they hold
on any bounded neighbourhood
$U\subset C$
of an arbitrary point
$(t,x)\in C$.
We can assume without loss of generality 
that the boundary
$\partial U$
is smooth.
Then the parabolic boundary value problem for
$g:\overline{U}\rightarrow\RR$,
given by
$\L g = 0$
in
$U$
and
$g\arrowvert_{\partial U} =Z\arrowvert_{\partial U}$,
has a unique solution (see~\cite{Friedman}, Theorem 6.3.6).
Let
$\tau_U$
be the first time the process
$Y$,
which started 
at
$(t,x)\in U$,
hits
$\partial U$.
From Dynkin's formula (see~\cite{Oksendal},
Theorem 7.4.1) and the fact that 
$g$
is the solution of the above Dirichlet problem
we find
$g(t,x)=\EE_{t,x}[Z(Y_{t+\tau_U})]$.
Since 
$Z$
satisfies the mean-value property (see~\cite{Oksendal},
page 121, formula (7.2.9)),
it follows that 
$g(t,x)=Z(t,x)$
for all
$(t,x)\in U$.
This proves the theorem in the double-barrier case. 

If we only have a lower barrier, 
we can express the function 
$Z$
as a limit of double-barrier option prices
where the ``artificial'' upper barrier 
tends to 
infinity.
Using a similar argument as above, and the fact that 
the maximum of the process
$X$
is finite
$\QQ$-almost surely in the time interval
$[0,T]$
(by Lemma~\ref{lem:Boundary} the process 
$X$
does not explode to infinity in finite time)
it is not hard to see that the convergence is locally
uniform, which in turn implies that the function 
$Z$
is continuous on the complement (in
$\overline{C}$)
of the discontinuities of
$\phi$.
Once we have established continuity,
the same ``local'' argument as in the paragraph above
proves the theorem in the case where there is no upper barrier.

In the up-and-out case, we introduce a constant lower barrier at some 
small level
$\epsilon$
with the boundary condition
$Z(t,\epsilon)=\phi(\epsilon)$
for all
$t\in[0,T]$.
Since the function 
$\phi$
is continuous at zero, the argument similar to the one above
yields continuity of the solution of the parabolic problem  obtained
in the limit as
$\epsilon\rightarrow0$.
Once we have continuity of the solution, the ``local'' behaviour 
follows as in the preceding two cases.
\end{pr}

%% file: proof.tex
We start by proving Lemma~\ref{lem:Boundary}.

\begin{pr}
By Feller's test for explosions (see Theorem~5.5.29 in~\cite{Karatzas})
it is well know that the statement of the lemma holds if the following
iterated integral diverges
$$\int_z^\infty\frac{dx}{x^2\sigma(x)^2}\exp(B(x))\int_x^\infty\exp(-B(y)) dy=\infty,
\quad\mathrm{where}\quad B(x):=2\mu\int_{x_0}^x\frac{du}{u\sigma(u)^2}, $$
for some
$z,x_0\in(0,\infty)$.
This is clearly true if the limit
$\lim_{x\rightarrow\infty}B(x)$
is finite.
If this is not the case, a simple application of L'H\^{o}pital's rule
implies that the function
$x\mapsto\exp(B(x))\int_x^\infty\exp(-B(y))dy$
is asymptotically equal to the function
$x\mapsto x\sigma(x)^2$.
This proves the lemma.
\end{pr}

Next is the proof of Lemma~\ref{lem:regularD}. 

\begin{pr}
The lemma is a consequence of Schauder's boundary estimates for the
solutions of the initial parabolic partial differential equations
proved in~\cite{Friedman1}.
Let us first consider the double-barrier case. 
Recall that 
$C$
denotes the domain of the solution of the PDE
from 
(b) of Lemma~\ref{lem:Z_Sub}.
Denote by
$F:\partial C-(\{0\}\times (b_-(0),b_+(0)))\rightarrow\RR_+$
a continuous function which maps the curves
$b_\pm([0,T])$
to zero and coincides with the payoff 
$\phi$
on the interval
$[b_-(T),b_+(T)]$.
By Theorem 3.3.7 on page 65
in~\cite{FriedmanPar}
the partial derivatives 
$Z_S, Z_{SS}, Z_t$
of the solution 
$Z$
of the PDE in (b) of Lemma~\ref{lem:Z_Sub}
will be H\"older continuous of order
$\alpha\in(0,1)$
on 
$C$, 
if we can find an extension 
$\Psi:\overline{C}\rightarrow\RR_+$
of the
function
$F$,
whose partial derivatives 
$\Psi_S, \Psi_{SS}, \Psi_t$
are 
H\"older continuous of order
$\alpha$
on the domain
$C$.
Theorem 3.3.7
in~\cite{FriedmanPar} applies in our case
because the volatility function
$x\mapsto x\sigma(x)$
is
uniformly elliptic on the domain
$C$
since it is strictly positive on the  compact set
$\overline{C}$.
Before constructing an extension 
$\Psi$
explicitly, let us show that H\"older continuity
of the partial derivatives of 
$Z$
implies the lemma.

Pick a sequence 
$(\epsilon_n)_{n\in\NN}$
of positive real numbers which converges to zero. Since the second derivative
$Z_{SS}$
is H\"older continuous on a bounded domain
$C$,
its modulus must be bounded by some constant
$c$
(i.e.
$|Z_{SS}(t,x)|<c$
for all points
$(t,x)\in C$).
Therefore, by Lagrange's theorem,
we have 
$|Z_S(t,b_+(t)-\epsilon_n) -Z_S(t, b_+(t)-\epsilon_k)|<c |\epsilon_n-\epsilon_k|$
for all
$n,k\in\NN$
and
all
$t\in[0,T]$.
Since the right-hand side of this inequality is independent of time
$t$,
the sequence of functions 
$(t\mapsto Z_S(t,b_+(t)-\epsilon_n))_{n\in\NN}$ 
is uniformly Cauchy on the interval
$[0,T]$
and therefore converges uniformly to the continuous limit
$\Delta_+$.
The same argument can be used for the lower barrier. This implies part
(a) of the lemma.

For part (b) let us choose a real number 
$\delta>0$
such that the point
$(t,b_+(t)-2\delta)$
lies in the domain
$C$
for all 
$t\in [0,T]$.
Since the barrier
$b_+$
is uniformly continuous on the interval
$[0,T]$,
there exists
$\delta_0>0$
with the following property: if
$|t-s|<\delta_0$,
then
$|b_+(t)-b_+(s)|<\delta$
for all
$s,t\in[0,T]$.
Choose any 
$\epsilon\in(0,\delta)$
and assume that 
$t_i\in[0,T]$
satisfy
$0=t_0<t_1<\ldots<t_n=T$
and
$\max\{t_i-t_{i-1};i=1,\ldots,n\}<\delta_0$.
Note that this implies that, if
$b_+(t_i)\geq b_+(t_{i-1})$,
for any
$i\in\{1,\ldots,n-1\}$,
the point
$(t_i,b_+(t_{i-1})-\epsilon)$
lies in the domain
$C$.
Similarly if
$b_+(t_i)<b_+(t_{i-1})$
we find that the point
$(t_{i-1},b_+(t_{i})-\epsilon)$
is in
$C$.
Using these observations we obtain
\begin{eqnarray*}
\lvert Z(t_i,b_+(t_{i})-\epsilon)-Z(t_{i-1},b_+(t_{i-1})-\epsilon)\rvert & \leq &
\lvert Z(t_i,b_+(t_{i})-\epsilon)-Z(t_{i-1},b_+(t_{i})-\epsilon)\rvert + \\
& & \lvert Z(t_{i-1},b_+(t_{i})-\epsilon) - Z(t_{i-1},b_+(t_{i-1})-\epsilon)\rvert \\
& \leq & A(t_i-t_{i-1}) + D \lvert b_+(t_i)-b_+(t_{i-1})\rvert \\
& \leq & (A+DE)(t_i-t_{i-1}), 
\end{eqnarray*}
where the constants
$A, D, E$
are upper bounds on the absolute values of the derivatives
$Z_t, Z_S, b_+'$
respectively.
In this inequality we assumed that 
$b_+(t_i)<b_+(t_{i-1})$.
In case 
$b_+(t_i)\geq b_+(t_{i-1})$
a similar bound with the same constants, which is also independent of
$\epsilon$,
can be obtained.
This inequality implies that the family of functions
$(t\mapsto Z(t,b_+(t)-\epsilon))_{\epsilon\in(0,\delta)}$
has a uniformly bounded total variation. 
Lower barrier can be dealt with in an analogous way.
This proves part (b) of our lemma.

We are now left with the task of showing that 
the payoff
$\phi$
can be extended to a function
$\Psi:\overline{C}\rightarrow\RR_+$
with H\"older continuous derivatives
$\Psi_t, \Psi_S, \Psi_{SS}$.
We start by defining a global diffeomorphism 
$\beta:[0,T]\times\RR\rightarrow [0,T]\times\RR$,
which straightens the barriers of the region 
$C$,
given by
$$\beta(t,S):=(t,B(t,S)),\>\>\>\>\>\>\>\>
\mathrm{where}\>\>\>\>\>\>\>\>B(t,S):=\frac{b_+(T)-b_-(T)}{b_+(t)-b_-(t)} (S-b_-(t))+b_-(T).$$
Note that 
$\beta(t,b_\pm(t))=(t,b_\pm(T))$
for all
$t\in[0,T]$
and
$\beta(T,S)=(T,S)$
for all
$S\in\RR$.
We can therefore define 
$\Psi(t,S):=\phi(B(t,S))$
for any point
$(t,S)\in \overline{C}$.
A simple calculation shows the following:
\begin{eqnarray*}
\Psi_S(t,S) & = & \phi'(B(t,S))\frac{b_+(T)-b_-(T)}{b_+(t)-b_-(t)}, \\
\Psi_{SS}(t,S) & = &  \phi''(B(t,S))\left(\frac{b_+(T)-b_-(T)}{b_+(t)-b_-(t)}\right)^2,\\
\Psi_t(t,S)  & = & - \phi'(B(t,S))\frac{b_+(T)-b_-(T)}{b_+(t)-b_-(t)}\left(b_-'(t)+
                     \frac{(S-b_-(t))(b_+'(t)-b_-'(t))}{b_+(t)-b_-(t)}\right). 
\end{eqnarray*}
The desired properties of the function
$\Psi$
follow directly from the assumptions in the lemma on
the payoff 
$\phi$
and the boundary functions
$b_\pm$.

Our final task is to prove the lemma in the case where there is only one barrier.
Theorem 3.3.7 on page 65
in~\cite{FriedmanPar}
can only be applied if the domain 
$C$
is bounded.
Assume we only have, say, a lower barrier
$t\mapsto b(t)$. 
Then by Theorem~\ref{thm:analyticity}
the discounted time-dependent single-barrier option price 
$Z(t,S)$
still solves the PDE from (b)
of Lemma~\ref{lem:Z_Sub}.
We can now introduce artificially a constant upper barrier at some
large value 
$B$
and
formulate a parabolic initial-boundary value problem
$U_t(t,S)+ \mu S U_S(t,S)+ \frac{S^2\sigma^2(S)}{2}U_{SS}(t,S)=0$
on the bounded domain
$C':=\{(t,S):t\in[0,T], S\in[b(t),B]\}$
with the payoff function
$\phi:[b(T),B]\rightarrow\RR$
and boundary conditions
$U(t,b(t))=0, U(t,B)=Z(t,B)$
for all
$t\in[0,T]$.
Like in the double-barrier case, because the domain
$C'$
is bounded, our assumption on the volatility function
$\sigma$
implies that the differential operator is uniformly elliptic.
By Theorem 4 in~\cite{Friedman1} such a problem has a unique
solution and therefore 
$U(t,S)=Z(t,S)$
for all
$(t,S)\in C'$.
Furthermore the same argument as above
implies that functions
$t\mapsto U_S(t,b(t)+\epsilon)$
converge uniformly to a continuous function
$t\mapsto \Delta(t)$
defined on
$[0,T]$
and that the total variation of
$t\mapsto U(t,b(t)+\epsilon)$
is bounded uniformly for all small positive 
$\epsilon$.
This proves the lemma in the case of a single-barrier option 
with a lower barrier.
The single upper barrier case can be dealt with similarly.
\end{pr}

Finally we demonstrate Lemma~\ref{lem:q_Reg}.

\begin{pr}
Since the set
$K$
is compact, it follows from definition~\eqref{eq:stock} that it
is enough to prove the lemma for the transition
density
$p(u-t;x,y)$
of the process
$S$.
As mentioned on page~\pageref{page:PDEq}
(see also~\cite{ItoMcKean},
page 149) the function
$(u,x)\rightarrow p(u-t;x,y)$
solves the parabolic PDE problem on the bounded
domain
$(t,T]\times K$.
Let the function
$(u,x)\rightarrow v(u-t;x,y)$
be the  solution of the same PDE satisfying
the boundary conditions
$v(u-t;x,y)=-p(u-t;x,y)$
for all
$u\in[t,T]$
and
$x\in\partial K$
(the symbol
$\partial K$
denotes the two boundary points of the interval
$K$)
and
the initial condition
$v(0;x,y)=0$
for all
$x\in K-\partial K$.
Such a solution exists and is bounded because the PDE is uniformly parabolic
on the domain
$(t,T]\times K$
and the boundary conditions are continuous and bounded.
Furthermore, by Section~5 in~\cite{FriedmanPar},
there exists a non-negative fundamental solution
$f$
for our parabolic PDE that satisfies the inequality in
Lemma~\ref{lem:q_Reg}.
By the maximum principle
(see~\cite{FriedmanPar}, Theorem 2.1 on page 34),
the function
$f$
dominates the solution
$v+p$
of the PDE
on the entire domain
$(t,T]\times K$.
Since the
$v$
is bounded the lemma follows.
\end{pr}